\documentclass[10pt, two column, twoside]{IEEEtran}

\usepackage{amssymb}
\usepackage{amsmath}
\usepackage{bm}
\usepackage[lined,boxed,commentsnumbered, ruled]{algorithm2e}
\usepackage{mathrsfs}
\usepackage{algorithmic}
\usepackage{tikz}
\usetikzlibrary{arrows}
\usepackage{subfigure}
\usepackage{graphicx,booktabs,multirow}

\definecolor{colorhkust}{RGB}{20,43,140}
\definecolor{colortsinghua}{RGB}{116,52,129}
\definecolor{color1}{RGB}{128,0,0}

\newtheorem{lemma}{Lemma}
\newtheorem{theorem}{Theorem}

\newtheorem{proposition}{Proposition}

\newtheorem{remark}{Remark}

\newcommand{\bs}{\boldsymbol}
\newcommand{\tabincell}[2]{\begin{tabular}{@{}#1@{}}#2\end{tabular}}
\newcommand{\rev}{\color{black}}
\newcommand{\RYS}[1]{\textcolor{black}{#1}}

\begin{document}

\title{Enhanced Group Sparse Beamforming for  Green Cloud-RAN: A Random
Matrix Approach}

\author{Yuanming~Shi,~\IEEEmembership{Member,~IEEE,}
 Jun~Zhang,~\IEEEmembership{Senior~Member,~IEEE,}
  Wei~Chen,~\IEEEmembership{Senior~Member,~IEEE,}
 and~Khaled~B.~Letaief,~\IEEEmembership{Fellow,~IEEE}
 \thanks{Y. Shi is with the School of Information Science and Technology,
ShanghaiTech University, Shanghai, China (e-mail: shiym@shanghaitech.edu.cn).}
\thanks{ J. Zhang is with the Department of Electronic and Computer Engineering,
 Hong Kong University of Science and Technology, Hong Kong (e-mail: eejzhang@ust.hk).}
\thanks{W. Chen is with the Department of Electronic Engineering, Tsinghua
University, Beijing, China (e-mail: wchen@tsinghua.edu.cn).}
\thanks{K. B. Letaief is with Hamad bin Khalifa University (e-mail: kletaief@hbku.edu.qa)
and Hong Kong University of Science and Technology (e-mail: eekhaled@ust.hk).}
\thanks{Part of this work was presented at the 2016 IEEE International Symposium on Information Theory (ISIT) \cite{Yuanming_ISIT16SGSBF}, Barcelona, Spain, Jul. 2016. This work was partly supported by the Hong Kong Research Grant Council under Grant No. 16200214.}}

\maketitle

\begin{abstract}
Group sparse beamforming is a general framework to minimize the network power consumption for cloud radio access networks (Cloud-RANs), which, however, suffers high computational complexity. In particular, a complex optimization problem needs to be solved to obtain the remote radio head (RRH) ordering criterion in each transmission block, which will help to determine the active RRHs and the associated fronthaul links. In this paper, we propose innovative approaches to reduce the complexity of this key step in group sparse beamforming. Specifically, we first develop a smoothed $\ell_p$-minimization approach with the  iterative reweighted-$\ell_2$ algorithm to {\RYS{return a Karush-Kuhn-Tucker (KKT) point solution, as well as}} enhancing the capability of inducing group sparsity in the beamforming vectors. By leveraging the Lagrangian duality theory, we obtain closed-form solutions at each iteration to reduce the computational complexity. The well-structured solutions provide the opportunities to apply the large-dimensional random matrix theory to derive deterministic approximations for the RRH ordering criterion. Such an approach helps to guide the RRH selection only based on the statistical channel state information (CSI), which does not require frequent update, thereby significantly
reducing the computation overhead. Simulation results shall demonstrate the performance gains of the proposed $\ell_p$-minimization approach, as well as the effectiveness  of the large system analysis based framework for computing RRH ordering criterion.         

\end{abstract}

\section{Introduction}
Network densification \cite{Bhushan_2014networkdensification, Yuanming_WCMLargeCVX, Andrews2015we} has been proposed as a promising way to provide ultra-high data rates, achieve low latency, and support ubiquitous connectivity for the upcoming 5G networks \cite{Jeff_JSAC5G}. However, to fully harness the benefits of dense wireless networks, formidable challenges arise, including interference management, radio resource allocation, mobility management, as well as high capital expenditure and operating expenditure. Cloud-RAN emerges as a disruptive technology to deploy cost-effective dense wireless networks \cite{Peng_HCRAN, Yuanming_TWC2014}. It can significantly improve both energy and spectral efficiency, by leveraging recent advances in cloud computing and network function virtualization \cite{Wubben_SPM2014Cloud}. With shared computation resources at the cloud data center and distributed low-cost low-power remote radio heads (RRHs), Cloud-RAN provides an ideal platform to achieve the benefits of network cooperation and coordination \cite{Gesbert_JSAC10}. There is a unique characteristic in Cloud-RANs, namely, the high-capacity fronthaul links are required to connect the cloud center and RRHs \cite{Peng_2015fronthaul, Shamai_SPM2014Fronthaul}. Such links will consume power comparable to that of each RRH, and thus brings new challenges to design green Cloud-RANs. To address this issue, a new performance metric, i.e., the \emph{network power consumption}, which consists of both the fronthaul link power consumption and RRH transmit power consumption, has been identified in  \cite{Yuanming_TWC2014} for the green Cloud-RAN design. 

Unfortunately, the network power minimization problem turns out to be a mixed-integer nonlinear programming problem \cite{leyffer_2012mixed}, which is highly intractable. Specifically, the combinatorial
composite objective function involves a \emph{discrete} component, indicating which RRHs and the corresponding fronthaul links should be switched on, and a \emph{continuous} component, i.e., beamformers to reduce RRH transmit power. To address this challenge, a unique group sparsity structure in the optimal beamforming vector has been identified in \cite{Yuanming_TWC2014} to unify  RRH selection and beamformer optimization. Accordingly, a novel three-stage \emph{group sparse beamforming} framework was proposed to promote group sparsity in the solution. Specifically, a mixed $\ell_1/\ell_2$-minimization approach was first proposed to induce group sparsity in the solution, thereby guiding the RRH selection via solving a sequence of convex feasibility problems, followed by coordinated beamforming to minimize the transmit power for the active RRHs.  

Although the group sparse beamforming framework provides polynomial-time complexity algorithms via convex optimization, it has the limited capability to enhance group sparsity compared to the non-convex approaches \cite{Boyd_2008enhancing, Daubechies_2010iteratively}. In particular, the smoothed $\ell_p$-minimization supported by the iterative reweighted-$\ell_2$ algorithm was developed in \cite{Yuanming_JSAC2015} to enhance sparsity for multicast group sparse beamforming. However, the computation burden of this iterative algorithm is still prohibitive in dense wireless networks with a large number of RRHs and mobile users. In \cite{Yuanming_LargeSOCP2014}, a generic two-stage approach was proposed to solve large-scale convex optimization problems in dense wireless cooperative networks via matrix stuffing and operator splitting, i.e., the alternating direction method of multipliers (ADMM) \cite{boyd2011distributed, Boyd_arXiv2013}. This approach also enables parallel computing and infeasibility detection. However, as the proposed solutions need to be accomplished for each channel realization, and also due to the iterative procedures of the group sparse beamforming framework, it is still computationally expensive.

 In this paper, we improve the performance of group sparse beamforming \cite{Yuanming_TWC2014}  via $\ell_p$-minimization, and a special emphasis is on reducing the computational complexity, with two ingredients: closed-form solutions for each iteration, and the deterministic equivalents of the optimal
Lagrange multipliers. Specifically, we first develop an iterative reweighted-$\ell_2$ algorithm
with closed-form solutions at each iteration via leveraging the principles of the majorization-minimization (MM) algorithm \cite{hunter2004tutorial} and the Lagrangian
duality
theory \cite{Yu_2007transmitter}. {\RYS{It turns out that the proposed iterative reweighted-$\ell_2$ algorithm can find a KKT point for the smoothed non-convex $\ell_p$-minimization problem.}} Furthermore, this reveals the explicit structures of the optimal
solutions to each subproblem in the iterative reweighted-$\ell_2$
algorithm, while the convex optimization approach in \cite{Wei_IA2014Sparse, Yuanming_TWC2014, Yuanming_JSAC2015}
fails to obtain the closed-form solutions. Thereafter, the well-structured closed-form solutions  provide opportunities to leverage the large-dimensional
random matrix theory \cite{Verdu_RMT04, Couillet_2011random, Debbah_TIT2012}
to perform the asymptotic analysis in the large system  regimes \cite{Hanly_2012TIT, Couillet_TWC16}. Specifically,
the deterministic equivalents of the optimal Lagrange multipliers are
derived based on the recent results of large random matrix analysis \cite{Couillet_2014large}
to decouple the dependency of system parameters. These results are further
used to perform an asymptotic analysis for computing the RRH ordering criterion, 
which only depends on long-term channel attenuation, thereby significantly reducing the computation overhead compared with previous algorithms that heavily depends on instantaneous CSI \cite{TonyQ.S._WC2013, Z.Q.Luo_JSAC2013, Rui_TWC2015GSBF, Wei_JSAC16, Yuanming_TWC2014, Wei_IA2014Sparse, Yuanming_JSAC2015}.    

Based on the above proposal, we provide a three-stage enhanced group sparse beamforming
framework for green Cloud-RAN  via random matrix theory. Specifically, in the first stage, we compute the enhanced RRH ordering criterion only based on the statistical CSI. This thus avoids frequent updates, thereby significantly reducing computational complexity, which is a sharp difference compared to the original
proposal in [8]. This new algorithm is based on the principles
of the MM algorithm and the Lagrangian duality theory, followed by the random matrix theory. With the obtained deterministic equivalents of RRH ordering criterion, a  two-stage large-scale convex
optimization framework \cite{Yuanming_LargeSOCP2014} is adopted to solve
a sequence of convex feasibility problems to determine the active RRHs in
the second stage, as well as solving the transmit power minimization problem
for the active RRHs in the final stage. Note that global instantaneous CSI
is required for these two stages. Simulation results are provided to demonstrate the improvement of the
enhanced group sparse beamforming framework. Moreover, the deterministic approximations
turn out to be accurate even in the finite-sized systems.

\subsection{Related Works}                  
\subsubsection{Sparse Optimization in Wireless Networks} The sparse optimization paradigm has recently been popular for complicated network optimization problems in wireless network design, e.g., the group sparse beamforming framework for green Cloud-RAN design {\cite{Yuanming_TWC2014, Rui_TWC2015GSBF, Wei_JSAC16, Yuanming_SDP2014, Yuanming_JSAC2015}}, wireless caching networks \cite{Jun_caching2014, Tao_TWC16}, user admission control \cite{Yuanming_JSAC2015, Tony_2015heterogeneous}, as well as computation offloading \cite{Yuanming_Compoffloading}. In particular, the convex relaxation approach provides a principled way to induce sparsity via the $\ell_1$-minimization \cite{Tony_2015heterogeneous} for individual sparsity inducing and the mixed $\ell_1/\ell_2$-minimization for group sparsity inducing \cite{Yuanming_TWC2014}. The reweighted-$\ell_1$ algorithm \cite{Boyd_2008enhancing} and the reweighted-$\ell_2$ algorithm \cite{Yuanming_JSAC2015} were further developed to enhance sparsity. To enable parallel and distributed computing, the first-order method ADMM algorithm was adopted to solve the group sparse beamforming problems \cite{Luo_2013base}. A generic large-scale convex optimization framework was further proposed to solve general large-scale convex programs in dense wireless networks to enable scalability, parallel computing and infeasibility detection \cite{Yuanming_LargeSOCP2014}. 

However, all the above algorithms need to be computed for each channel realization, which is computationally expensive. In this paper, we adopt large system analysis to compute the RRH ordering criterion for network adaptation only based on statistical CSI.          

\subsubsection{Large System Analysis via Random Matrix Theory} Random matrix theory \cite{Verdu_RMT04} has been proven powerful for performance analysis, and the understanding and improving algorithms in wireless communications \cite{Rusek_SPM2013, Couillet_2011random}, signal processing \cite{Couillet_SPM13LS}, and machine learning \cite{Tropp_15RMT, Couillet_2016randomICML}, especially in large dimensional regimes for applications in the era of big data \cite{Ding_CMagBigData}. In dense wireless networks, random matrix theory provides a powerful way for performance analysis and algorithm design. In particular, the large system analysis was performed for simple
precoding schemes, e.g., regularized zero-forcing in MISO broadcast channels
with imperfect CSI \cite{Debbah_TIT2012}. A random matrix approach to the optimal coordinated multicell beamforming for massive MIMO was presented in \cite{Debbah_MMIMOTIT2015} without close-form expressions for the optimal Lagrange multipliers. A simple channel model with two cells under different coordination levels was considered in \cite{Hanly_2012TIT}. 

However, all of the above results cannot be directly applied in the group sparse beamforming framework in dense Cloud-RAN due to the general channel models with heterogenous pathloss and the complicated beamformer structures with fully cooperative transmission. To determine the RRH ordering criterion based on statistical CSI, we adopt the technique in \cite{Couillet_2014large, Couillet_TWC16} to compute the closed-forms for the optimal Lagrange multipliers for each iteration in the procedure of reweighted-$\ell_2$ minimization for sparsity inducing.       

\subsection{Organization}
The remainder of the paper is organized as follows. Section {\ref{sys}} presents the system model and problem formulation, followed by performance analysis with the proposed three-stage enhanced group sparse beamforming framework for green Cloud-RAN. In Section {\ref{rewgsbf}}, the iterative reweighted-$\ell_2$ algorithm for group sparse beamforming is presented via the principle of MM algorithm and duality theory. The large system analysis for RRH ordering is performed in Section {\ref{lsa}}. Simulation results will be demonstrated in Section {\ref{simres}}. Finally, conclusions and discussions are presented in Section {\ref{condis}}. To keep the main text clean and free of technical details, we divert most of the proofs to the Appendix.

\subsubsection*{Notations} Throughout this paper, $\|\cdot\|_p$ is the $\ell_p$-norm.  $|\cdot|$ stands for either the size
of a set or the absolute value of a scalar, depending on the context. $\|\bm{M}\|$ is the spectral radius of the Hermitian matrix $\bm{M}$. Boldface lower case and upper case letters represent vectors and matrices, respectively. $(\cdot)^{-1}, (\cdot)^T, (\cdot)^{\sf{H}}$ and ${\rm{Tr}}(\cdot)$ denote the inverse, transpose, Hermitian and trace operators, respectively. We use $\mathbb{C}$ to represent complex domain. $\mathbb{E}[\cdot]$ denotes the expectation of a random variable.  We denote ${\bf{A}}={\rm{diag}}\{x_1,\dots, x_N\}$ and ${\bf{I}}_N$ as a diagonal matrix of order $N$ and the identity matrix of order $N$, respectively.

\section{System Model and Problem Formulation}
\label{sys}
\subsection{System Model}
Consider a Cloud-RAN with $L$ RRHs and $K$ single-antenna mobile users (MUs), where each RRH is equipped with $N$ antennas.  In Cloud-RANs, the BBU pool will perform the centralized signal processing and is connected to all the RRHs via high-capacity and low-latency fronthaul links. In this paper, we will focus on the downlink signal processing. 
Specifically, let ${\bf{v}}_{lk}\in\mathbb{C}^{N}$ be the transmit beamforming vector
from the $l$-th RRH to the $k$-th MU. The received signal $y_{k}\in\mathbb{C}$ at MU $k$ is  given by
\setlength\arraycolsep{1.2pt}
\begin{eqnarray}
y_{k}=\sum_{l=1}^{L}{\bf{h}}_{kl}^{\sf{H}}{\bf{v}}_{lk}s_{k}+\sum_{i\ne
k}\sum_{l=1}^{L}{\bf{h}}_{kl}^{\sf{H}}{\bf{v}}_{li}s_{i}+n_{k}, 
\end{eqnarray} 
where ${\bf{h}}_{kl}\in\mathbb{C}^{N}$ is the channel
propagation between  MU $k$ and RRH $l$, $s_k\in\mathbb{C}$
with $\mathbb{E}[|s_k|^2]=1$ is the encoded transmission information symbol
for MU $k$, and $n_{k}\sim\mathcal{CN}(0, \sigma_{k}^2)$ is the additive
Gaussian noise at MU $k$.

We assume that $s_{k}$'s and $n_{k}$'s are mutually independent and all the MUs apply single-user detection. The signal-to-interference-plus-noise ratio (SINR) for MU $k$ is given by
\begin{eqnarray}
\label{sinrexp}
{{\sf{sinr}}}_{k}={{|{\bf{h}}_{k}^{\sf{H}}{\bf{v}}_{k}|^2}\over{\sum_{i\ne
k}|{\bf{h}}_{k}^{\sf{H}}{\bf{v}}_{i}|^2+\sigma_{k}^2}},
\end{eqnarray}
where ${\bf{h}}_{k}= [{\bf{h}}_{k1}^{T},\dots, {\bf{h}}_{kL}^{T}]^{T}\in\mathbb{C}^{LN}$ is the vector consisting of the channel coefficients  from all the RRHs to MU $k$, and ${\bf{v}}_{k}=[{\bf{v}}_{1k}^{T}, \dots, {\bf{v}}_{Lk}^{T}]^{T}\in\mathbb{C}^{LN}$ is the
aggregative beamforming vector for the MU $k$ from all the RRHs.

\subsection{Problem Formulation}
In this paper, we aim at designing a green Cloud-RAN by minimizing the network power consumption, which consists of the fronthaul links power consumption, as well as the RRH transmit power consumption \cite{Yuanming_TWC2014}. Specifically, let ${\bf{v}}=[{\bf{v}}_{lk}]\in\mathbb{C}^{KLN}$ be the aggregative beamforming vector from all the RRHs to all the MUs. Define the support of the beamforming vector $\bf{v}$ as $\mathcal{T}({\bf{v}})=\{i|v_{i}\ne0\}$,
where ${\bf{v}}=[v_i]$ is indexed by $i\in\mathcal{V}$ with $\mathcal{V}=\{1,\dots,
KLN\}$. The relative fronthaul network power consumption is given by
\begin{eqnarray}
f_1({\bf{v}})=\sum\limits_{l=1}^{L}P_{l}^c I(\mathcal{T}({\bf{v}})\cap\mathcal{V}_{l}\ne\emptyset),
\end{eqnarray}
where $P_l^c\ge 0$ is the relative fronthaul link power consumption \cite{Yuanming_TWC2014} (i.e., the static power saving when both the fronthaul link and the corresponding RRH are switched off), $\mathcal{\mathcal{V}}_{l}=\{(l-1)KN+1,\dots,lKN\},
\forall l\in\mathcal{L}$, is a partition of $\mathcal{V}$, and $I(\mathcal{T}\cap\mathcal{V}_{l}\ne\emptyset)$ is an indicator function
that takes value 1 if $\mathcal{T}\cap\mathcal{V}_{l}\ne\emptyset$ and 0
otherwise. {\RYS{Note that  $f_1$ is a non-convex combinatorial function.}}  

Furthermore, the total transmit power consumption is given by  
\begin{eqnarray}
f_2({\bf{v}})=\sum_{l=1}^{L}\sum_{k=1}^{K}{1\over{\zeta_{l}}}\|{\bf{v}}_{lk}\|_2^2,
\end{eqnarray} 
where $\zeta_l>0$ is the drain inefficiency coefficient of the radio frequency power amplifier \cite{Yuanming_TWC2014}. Therefore, the network power consumption is represented by the  combinatorial composite function
\begin{eqnarray}
f({\bf{v}})=f_1({\bf{v}})+f_2({\bf{v}}).
\end{eqnarray}

Given the QoS thresholds ${\bs{\gamma}}=(\gamma_1,\dots, \gamma_K)$ for all the MUs, in this paper, we aim at solving the following network power consumption minimization problem with the QoS constraints:   
\begin{eqnarray}
\label{npm}
\mathscr{P}:
\mathop {\textrm{minimize}}_{{\bf{v}}}&& f_1({\bf{v}})+f_2({\bf{v}})\nonumber\\
\textrm{subject to}&&{{|{\bf{h}}_{k}^{\sf{H}}{\bf{v}}_{k}|^2}\over{\sum_{i\ne
k}|{\bf{h}}_{k}^{\sf{H}}{\bf{v}}_{i}|^2+\sigma_{k}^2}}\ge
\gamma_k, \forall k. 
\end{eqnarray}
Let $\tilde{\bf{v}}_l=[{\bf{v}}]_{\mathcal{V}_l}=[{\bf{v}}_{l1}^T,\dots, {\bf{v}}_{lK}^T]^T\in\mathbb{C}^{KN}$ form the beamforming coefficient group from RRH $l$ to all the MUs. Note that, when the RRH $l$ is switched off, all the beamforming coefficients in $\tilde{\bf{v}}_l$ will be set to zeros simultaneously. Observing that there may be multiple RRHs being switched off to minimize the network power consumption, the optimal beamforming vector ${\bf{v}}=[\tilde{\bf{v}}_1^T,\dots, \tilde{\bf{v}}_L^T]^T\in\mathbb{C}^{KLN}$ should have a group-sparsity structure. Therefore, problem $\mathscr{P}$ is called a \emph{group sparse beamforming problem} \cite{Yuanming_TWC2014}. {\RYS{Note that, to simplify the presentation, we only impose the QoS constraints
in problem $\mathscr{P}$. However, the proposed instantaneous CSI based iterative
reweighted-$\ell_2$ algorithm can be extended to the scenario with per-RRH
transmit power constraints, following the principles in \cite{Yu_2007transmitter}. More technical
efforts are required to derive the asymptotic results using the random matrix
method with more complicated structures of the optimal beamformers.}}

\subsection{Problem Analysis}
Although the constraints in problem $\mathscr{P}$ can be reformulated as convex second-order cone constraints, the non-convex objective function makes it highly intractable. To address this challenge, a weighted mixed $\ell_1/\ell_2$-norm minimization approach was proposed in \cite{Yuanming_TWC2014} to convexify the objective and induce the group sparsity in beamforming vector $\bf{v}$, thereby guiding the RRH ordering to enable adaptively RRH selection. Specifically, we will first solve the $\ell_1/\ell_2$-norm minimization problem, and denote $\tilde{\bf{v}}_1^{\star},\dots, \tilde{\bf{v}}_L^{\star}$ as the induced  (approximated) group sparse beamforming vectors. Then the following RRH ordering criterion is adopted to determine which
RRHs should be switched off \cite{Yuanming_TWC2014}:
\begin{eqnarray}
\label{rrhorder}
{\theta}_l={\kappa}_l\|\tilde{\bf{v}}_l^{\star}\|_2^2, \forall l=1,\dots, L,
\end{eqnarray}
where $\kappa_l=\sum_{k=1}^K\|{\bf{h}}_{kl}\|_2^2/\nu_l$. In particular,
the group sparsity structure information for beamforming vector $\bf{v}$
is extracted from the squared $\ell_2$-norm of the beamforming vectors $\tilde{\bf{v}}_l$'s,
i.e., $\|\tilde{\bf{v}}_1\|_2^2, \dots, \|\tilde{\bf{v}}_L\|_2^2$. The RRH with a smaller ${\theta}_l$ will have a higher priority to be switched off. Based
on the RRH ordering result $\bm{\theta}^{\star}$ in (\ref{rrhorder}), a bi-section search approach can be
used to find the optimal active RRHs  \cite{Yuanming_TWC2014} via solving a sequence of the following feasibility problems:
\begin{eqnarray}
\label{feasiblecheck}
\mathscr{F}(\mathcal{A}^{[i]}):\mathop {\textrm{find}}&&
{\bf{v}}_1,\dots, {\bf{v}}_K\nonumber\\
\textrm{subject to}&&{{|{\bf{h}}_{k}^{\sf{H}}{\bf{v}}_{k}|^2}\over{\sum_{i\ne
k}|{\bf{h}}_{k}^{\sf{H}}{\bf{v}}_{i}|^2+\sigma_{k}^2}}\ge
\gamma_k, \forall k, 
\end{eqnarray}
where ${\bf{v}}_k=[{\bf{v}}_{lk}]\in{\mathbb{C}}^{|\mathcal{A}|N}$ and ${\bf{h}}_k=[{\bf{h}}_{kl}]\in\mathbb{C}^{|\mathcal{A}|N}$. Problem $\mathscr{F}(\mathcal{A}^{[i]})$ turns out to be convex via reformulating the QoS constraints as second-order cone constraints \cite{Yuanming_TWC2014}. To further enhance the sparsity as well as to seek the quadratic
forms of the beamforming vectors in the multicast transmission setting, a smoothed $\ell_p$-minimization approach was  proposed in \cite{Yuanming_JSAC2015}. To scale to large problem sizes in dense Cloud-RANs, a two-stage large-scale parallel convex optimization framework was developed in \cite{Yuanming_LargeSOCP2014} with the capability of infeasibility detection.

However, all of the above developed algorithms bear high computation overhead. In particular, while the feasibility problem (8) for RRH selection can be efficiently solved with the large-scale optimization algorithm in [8], the RRH ordering criterion in (\ref{rrhorder}) may be highly complicated to obtain, especially with the non-convex formulation as in \cite{Yuanming_JSAC2015}. Moreover, the ordering criterion needs to be recomputed for each transmission slot, and depends on instantaneous CSI. Observing that the statistical CSI normally changes much slower than the instantaneous CSI, to reduce the computational burden, we propose to compute the RRH ordering criterion (\ref{rrhorder}) only based on statistical CSI, i.e., the long-term channel attenuation. This is achieved by first developing a group sparsity penalty with quadratic forms in the aggregative beamforming vector ${\bf{v}}$, followed by an iterative reweighted-$\ell_2$ algorithm with closed-form solutions at each iteration via Lagrangian duality theory, as will be presented in Section \ref{rewgsbf}. Then  asymptotic analysis is performed to obtain the RRH ordering criterion (\ref{rrhorder}) based only on statistical CSI by leveraging the large-dimensional random matrix theory \cite{Couillet_2011random, Debbah_TIT2012,Couillet_2014large}, as will be presented in Section \ref{lsa}. 

\begin{figure}[t]
  \centering
  \includegraphics[width=0.9\columnwidth]{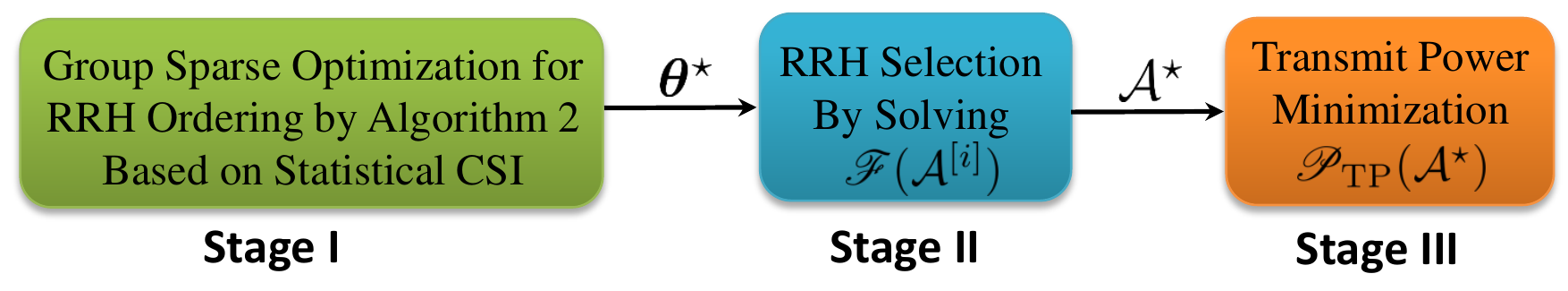}
 \caption{The proposed three-stage enhanced group sparse beamforming framework for
dense green Cloud-RAN. In the first stage, the RRH ordering criterion $\bm{\theta}$
is computed only based on the statistical CSI via large system analysis.
The optimal active RRHs $\mathcal{A}^{\star}$ in the second stage and the
optimal coordinated beamforming for transmit power minimization in the third
stage are computed based on the instantaneous CSI via the large-scale convex
optimization algorithm in \cite{Yuanming_LargeSOCP2014}.}
 \label{GSBF}
\end{figure}

Overall, the proposed three-stage enhanced group sparse beamforming framework is presented in Fig. {\ref{GSBF}}. Specifically, in the first stage, the RRH ordering criterion $\bm{\theta}^{\star}$ is calculated only based on statistical CSI using Algorithm \ref{irm1sta}. In the second stage, the set of active RRHs $\mathcal{A}^{\star}$ is obtained based on  instantaneous CSI via solving a sequence of convex feasibility problems $\mathscr{F}(\mathcal{A}^{[i]})$ using the large-scale convex optimization framework in \cite{Yuanming_LargeSOCP2014}. In the third stage, the transmit power is minimized by solving the convex program (\ref{npm}) with the fixed active RRHs $\mathcal{A}^{\star}$ using the large-scale convex optimization algorithm in \cite{Yuanming_LargeSOCP2014}. Overall, the proposed enhanced group sparse beamforming framework is scalable to large network sizes. This paper will focus on developing an effective algorithm for the first stage.

\begin{remark}
\label{reminfea} 
{\rev{In this paper, we assume that, in the first stage, problem $\mathscr{P}$ ({\ref{npm}}) is feasible for developing Algorithm {\ref{irm1}} and Algorithm {\ref{irm1sta}} to determine the RRH ordering criterion based on the instantaneous CSI and statistical CSI, respectively. {\RYS{We enable the capability of handing the infeasibility in the second stage based on the instantaneous CSI, using the large-scale convex optimization algorithm  \cite{Yuanming_LargeSOCP2014}. Furthermore, when problem $\mathscr{F}$ (\ref{feasiblecheck}) is infeasible with all  RRHs active, we adopt the user admission algorithm to find the maximum number of admitted users \cite{Yuanming_JSAC2015}.}}}}
\end{remark}

\section{Group Sparse Beamforming via Iterative Reweighted-$\ell_2$ Algorithm}
\label{rewgsbf}
In this section, we develop a group sparse beamforming approach based on the smoothed $\ell_p$-minimization, supported by an iterative reweighted-$\ell_2$ algorithm, thereby enhancing the group sparsity in the beamforming vectors. Instead of reformulating the QoS constraints in problem $\mathscr{P}$ as second-order cone constraints \cite{Yuanming_TWC2014}, we use the Lagrangian duality theory to reveal the structures of the  optimal solutions  at each iteration. The results will assist the large system analysis in Section \ref{lsa}.  

\subsection{Group Sparsity Inducing Optimization via Smoothed $\ell_p$-Minimization} To induce the group sparsity structure in the beamforming vector $\bf{v}$, thereby guiding the RRH ordering, the weighted mixed $\ell_1/\ell_2$-norm was proposed in \cite{Yuanming_TWC2014}. However, the non-smooth mixed $\ell_1/\ell_2$-norm fails to introduce the quadratic forms in the beamforming vector $\bf{v}$, in order to be compatible with the quadratic QoS constraints so that the Lagrangian duality theory can be applied \cite{Yu_2007transmitter}. To address this challenge, we adopt the following smoothed $\ell_p$-minimization ($0<p\le 1$) approach to induce group sparsity \cite{Daubechies_2010iteratively, Yuanming_JSAC2015}:
\begin{eqnarray}
\label{slp}
\mathscr{P}_{\textrm{GSBF}}: \mathop {\textrm{minimize}}_{{\bf{v}}}&&
g_p({\bf{v}}; \epsilon):=\sum_{l=1}^{L}\nu_l\left(\|\tilde{\bf{v}}_l\|_2^2+\epsilon^2\right)^{p/2}
 \nonumber\\
\textrm{subject to}&&{{|{\bf{h}}_{k}^{\sf{H}}{\bf{v}}_{k}|^2}\over{\sum_{i\ne
k}|{\bf{h}}_{k}^{\sf{H}}{\bf{v}}_{i}|^2+\sigma_{k}^2}}\ge
\gamma_k, \forall k,
\end{eqnarray}
where $\epsilon>0$ is some fixed regularizing parameter and $\nu_l>0$ is the weight for the beamforming coefficient group $\tilde{\bf{v}}_l$ by encoding the prior information of system parameters \cite{Yuanming_TWC2014}. Compared with the mixed $\ell_1/\ell_2$-norm
minimization approach, the $\ell_p$-minimization approach can induce sparser
solutions based on the fact  $\|{\bs{z}}\|_0=\lim_{p\rightarrow 0}\|{\bs{z}}\|_p^p=\lim_{p\rightarrow
0}\sum_i |z_i|^p$. Unfortunately, problem (\ref{slp}) is non-convex due to the non-convexity of both the objective function and the QoS constraints.

\subsection{Iterative Reweighted-$\ell_2$ Algorithm}
We use the MM algorithm and the Lagrangian duality theory to solve problem (\ref{slp}) with closed-form solutions. Specifically, this approach generates the iterates $\{{\bf{v}}^{[n]}\}$ by successively minimizing upper bounds $Q({\bf{v}}; {\bf{v}}^{[n]})$ of the objective function $g_p({\bf{v}};\epsilon)$. We adopt the following upper bounds to approximate the smoothed $\ell_p$-norm $g({\bf{v}}; \epsilon)$ by leveraging the results of the expectation-maximization (EM) algorithm \cite{Lange_1993normal}.

\begin{proposition}Given the value of ${\bf{v}}^{[n]}$ at the $n$-th iteration, an upper bound for the objective function $g_p({\bf{v}}; \epsilon)$ can be constructed as follows: 
\begin{eqnarray}
Q({\bf{v}}; {\bs{\omega}}^{[n]}):=\sum_{l=1}^L\omega_l^{[n]}\|\tilde{\bf{v}}_l\|_2^2,
\end{eqnarray}
where 
\begin{eqnarray}
\label{upweight}
\omega_l^{[n]}= {{p\nu_l}\over{2}}\left[\left\|\tilde{\bf{v}}_l^{[n]}\right\|_2^2+\epsilon^2\right]^{{p\over{2}}-1}, \forall l=1,\dots, L.
\end{eqnarray}
\begin{IEEEproof}
{\rev{The proof is mostly based on  \cite[Proposition 1]{Yuanming_JSAC2015}.}} 
\end{IEEEproof}
\end{proposition}  

Therefore, at the $n$-th iteration, we need to solve the following optimization problem: 
\begin{eqnarray}
\mathscr{P}_{\textrm{GSBF}}^{[n]}: \mathop {\textrm{minimize}}_{{\bf{v}}}&&
\sum_{l=1}^{L}\omega_l^{[n]}\|\tilde{\bf{v}}_l\|_2^2
 \nonumber\\
\textrm{subject to}&&{{|{\bf{h}}_{k}^{\sf{H}}{\bf{v}}_{k}|^2}\over{\sum_{i\ne
k}|{\bf{h}}_{k}^{\sf{H}}{\bf{v}}_{i}|^2+\sigma_{k}^2}}\ge
\gamma_k, \forall k,
\end{eqnarray}
which is non-convex due to the non-convex QoS constraints. Although the QoS constraints can be reformulated as second-order cone constraints as in \cite{Yuanming_TWC2014}, in this paper, we leverage the Lagrangian duality theory to obtain explicit structures of the optimal solution to problem $\mathscr{P}_{\textrm{GSBF}}^{[n]}$, thereby reducing the computational complexity and further aiding the large system analysis in next section. {\RYS{The success of applying the Lagrangian duality approach is based on
the fact that strong duality holds for problem $\mathscr{P}_{\textrm{GSBF}}^{[n]}$
\cite{WeiYu_WC10}, i.e., the gap between the primal optimal objective and the dual optimal
objective is zero.}}

\subsubsection{Simple Solution Structures}
As the strong duality holds for problem $\mathscr{P}_{\textrm{GSBF}}^{[n]}$ \cite{WeiYu_WC10}, we solve it using duality theory. Specifically, let $\lambda_k/(NL)\ge 0$ denote the Lagrange multipliers corresponding to
the QoS constraints in problem $\mathscr{P}_{\textrm{GSBF}}^{[n]}$. Define the Lagrangian function $\mathcal{L}(\{{\bf{v}}_k\}, {\bs{\lambda}})$ with ${\bs{\lambda}}=\{\lambda_i\}$ as 
\begin{eqnarray}
\label{lagmul}
\mathcal{L}(\{{\bf{v}}_k\}, {\bs{\lambda}})&=&\sum_{k=1}^{K}{\lambda_k\over{LN}}\left(\sum_{i\ne
k}|{\bf{h}}_k^{\sf{H}}{\bf{v}}_i|^2+\sigma_k^2-{1\over{\gamma_k}}|{\bf{h}}_k^{\sf{H}}{\bf{v}}_k|^2\right)\nonumber\\
&&+\sum_{k=1}^{K}{\bf{v}}_k^{\sf{H}}{\bf{Q}}^{[n]}{\bf{v}}_k,
\end{eqnarray}
where ${\bf{Q}}^{[n]}\in\mathbb{C}^{LN\times LN}$ is a block diagonal matrix with the scaled identity matrix $\omega_l^{[n]}{\bf{I}}_N$ as the $l$-th main diagonal block square matrix.

To find the optimal ${\bf{v}}_k$'s, we take the gradient of the Lagrangian function $\mathcal{L}({\bf{v}}, {\bs{\lambda}})$ with respect to ${\bf{v}}_k$ and set it to zero, which implies
\begin{eqnarray}
\label{lagzero}
{\bf{Q}}^{[n]}{\bf{v}}_k+\sum_{i\ne k}{{\lambda_k}\over{LN}}{\bf{h}}_i{\bf{h}}_i^{\sf{H}}{\bf{v}}_k-{\lambda_k\over{LN\gamma_k}}{\bf{h}}_k{\bf{h}}_k^{\sf{H}}{\bf{v}}_k={\bf{0}}. \end{eqnarray}
By adding $\lambda_k/(LN)\times{\bf{h}}_k{\bf{h}}_k^{\sf{H}}{\bf{v}}_k$ to
both sides of (\ref{lagzero}), we have 
\begin{eqnarray}
\label{lagfinal}
\!\!\!\!\!\!\!\left({\bf{Q}}^{[n]}+\sum_{i=1}^{K}{{\lambda_i}\over{LN}}{\bf{h}}_i{\bf{h}}_i^{\sf{H}}\right){\bf{v}}_k={\lambda_k\over{LN}}\left(1+{1\over{\gamma_k}}\right){\bf{h}}_k{\bf{h}}_k^{\sf{H}}{\bf{v}}_k,
\end{eqnarray}
which implies ${\bf{v}}_k=\left({\bf{Q}}^{[n]}+\sum_{i=1}^{K}{{\lambda_i}/({LN})}{\bf{h}}_i{\bf{h}}_i^{\sf{H}}\right)^{-1}{\bf{h}}_k\times ({\lambda_k/({LN})})\left(1+{1/{\gamma_k}}\right){\bf{h}}_k^{\sf{H}}{\bf{v}}_k$. As $({\lambda_k/{LN}})\left(1+{1/{\gamma_k}}\right){\bf{h}}_k^{\sf{H}}{\bf{v}}_k$ is a scalar, the optimal ${\bf{v}}_k$ must be parallel to the beamforming direction
\begin{eqnarray}
\label{bfs1}
\bar{\bf{v}}_k=\left({\bf{Q}}^{[n]}+\sum_{i=1}^{K}{{\lambda_i}\over{LN}}{\bf{h}}_i{\bf{h}}_i^{\sf{H}}\right)^{-1}{\bf{h}}_k,
\forall k.
\end{eqnarray}  
Therefore, the optimal beamforming vectors ${\bf{v}}_1, \dots, {\bf{v}}_K$ can be written as
\begin{eqnarray}
\label{optbf}
{\bf{v}}_k=\sqrt{p_k\over{LN}}{\bar{\bf{v}}_k\over{\|\bar{\bf{v}}_k\|_2}}, \forall k,
\end{eqnarray}
where $p_k$ denotes the optimal beamforming power. Observing that the beamforming powers $p_k$'s need to satisfy
the SINR constraints with equality \cite{Yu_2007transmitter}, i.e.,
\begin{eqnarray}
\label{bfs2}
{{p_k}\over{\gamma _kLN}}{{|{\bf{h}}_k^{\sf{H}}\bar{\bf{v}}_k|^2}\over{\|\bar{\bf{v}}_k\|_2^2}}-\sum_{i\ne
k}{p_i\over{LN}}{{|{\bf{h}}_k^{\sf{H}}\bar{\bf{v}}_i|^2}\over{\|\bar{\bf{v}}_i\|_2^2}}=\sigma_k^2,
\forall k,
\end{eqnarray} 
the optimal powers  thus can be obtained via solving the following linear equation:
\begin{eqnarray}
\label{dualpower}
\left[ \begin{array}{ccc}
p_1  \\
\vdots \\
p_K 
\end{array} \right]={\bf{M}}^{-1}\left[ \begin{array}{ccc}
\sigma_1^2  \\
\vdots \\
\sigma_K^2 
\end{array} \right],
\end{eqnarray}
where
\begin{displaymath}
[{\bf{M}}]_{ij} = \left\{ \begin{array}{ll}
{{1}\over{\gamma _iLN}}{{|{\bf{h}}_i^{\sf{H}}\bar{\bf{v}}_i|^2}\over{\|\bar{\bf{v}}_i\|_2^2}}
&, i= j,\\
-{1\over{LN}}{{|{\bf{h}}_i^{\sf{H}}\bar{\bf{v}}_j|^2}\over{\|\bar{\bf{v}}_j\|_2^2}}
&, i\ne j.
  \end{array} \right.
\end{displaymath}
Here, 
 $[{\bf{M}}]_{ij}$ denotes the $(i,j)$-th element of the matrix ${\bf{M}}\in\mathbb{R}^{K\times
K}$. 

To find the optimal $\bs{\lambda}$-parameter, by multiplying both sides by ${\bf{h}}_k^{\sf{H}}\left({\bf{Q}}^{[n]}+\sum_{i=1}^{K}{{\lambda_i}\over{LN}}{\bf{h}}_i{\bf{h}}_i^{\sf{H}}\right)^{-1}$
in (\ref{lagfinal}), we obtain ${\bf{h}}_k^{\sf{H}}{\bf{v}}_k$ as
\begin{eqnarray}
\label{lambdeq}
{\lambda_k\over{LN}}\left(1+{1\over{\gamma_k}}\right){\bf{h}}_k^{\sf{H}}\left({\bf{Q}}^{[n]}+\sum_{i=1}^{K}{{\lambda_i}\over{LN}}{\bf{h}}_i{\bf{h}}_i^{\sf{H}}\right)^{-1}\!\!\!\!\!\!\!{\bf{h}}_k{\bf{h}}_k^{\sf{H}}{\bf{v}}_k,
\end{eqnarray}
which implies 
\begin{eqnarray}
\label{dualpara}
\!\!\!\!\!\lambda_k=LN\left[\left(1+{1\over{\gamma_k}}\right){\bf{h}}_k^{\sf{H}}\left({\bf{Q}}^{[n]}+\sum_{i=1}^{K}{{\lambda_i}\over{LN}}{\bf{h}}_i{\bf{h}}_i^{\sf{H}}\right)^{-1}\!\!\!\!\!\!\!{\bf{h}}_k\right]^{-1}\!\!\!\!\!\!\!\!.
\end{eqnarray} 
These fixed-point equations can be computed using iterative function
evaluation. 

Based on (\ref{optbf}), (\ref{dualpower})
and (\ref{dualpara}), we obtain the squared $\ell_2$-norm of the optimal solution to problem $\mathscr{P}_{\textrm{GSBF}}^{[n]}$ as follows:
\begin{eqnarray}
\label{sel2}
\left\|\tilde{\bf{v}}_l^{[n]}\right\|_2^2=\sum_{k=1}^{K}{\bf{v}}_k^{\sf{H}}{\bf{Q}}_{lk}{\bf{v}}_k,
\end{eqnarray}
where ${\bf{Q}}_{lk}\in\mathbb{C}^{NL\times NL}$ is a block diagonal matrix
with the identity matrix ${\bf{I}}_N$ as the $l$-th main diagonal block square
matrix and zeros elsewhere.  Note that the optimal powers $p_k$'s (\ref{dualpower}) and  Lagrangian multipliers $\lambda_k$'s (\ref{dualpara}) should depend on the weights $\omega_l^{[n]}$'s (\ref{upweight}) at each iteration $n$.

The instantaneous CSI based iterative
reweighted-$\ell_2$ algorithm for group sparse beamforming is presented
in Algorithm {\ref{irm1}}. We have the following result for its performance and convergence. 
\begin{algorithm}
\label{irm1}
\caption{Instantaneous CSI based Iterative Reweighted-$\ell_2$ Algorithm for Problem $\mathscr{P}_{\textrm{GSBF}}$}
{\textbf{input}}: Initialize ${\bs{\omega}}^{[0]}=(1, \dots, 1)$;
$I$ (the maximum number of iterations)\\
 Repeat
 
~~1) Compute the squared $\ell_2$-norm of the solution to problem $\mathscr{P}_{\textrm{GSBF}}^{[n]}$, $\|\tilde{\bf{v}}_1^{[n]}\|_2^2, \dots, \|\tilde{\bf{v}}_L^{[n]}\|_2^2$, using (\ref{optbf}), (\ref{dualpower}), (\ref{dualpara}) and (\ref{sel2}). 

~~2) Update the weights $\omega_l^{[n+1]}$ using (\ref{upweight}). 

Until convergence or attain the  maximum iterations and return {\textbf{output}}.\\
 {\textbf{output}}: RRH ordering criterion $\bm{\theta}$ (\ref{rrhorder}).
\label{gsbfal} 
\end{algorithm}

\begin{theorem}\label{tconvergence}
Let $\{{\bf{v}}^{[n]}\}_{n=1}^{\infty}$ be the sequence generated by the iterative reweighted-$\ell_2$ algorithm. Then, every limit point $\bar{\bf{v}}$ of $\{{\bf{v}}^{[n]}\}_{n=1}^{\infty}$ has the following properties:
\begin{enumerate}
\item $\bar{\bf{v}}$ is a KKT point of problem $\mathscr{P}_{\textrm{GSBF}}$ (\ref{slp});
\item $g_p({\bf{v}}^{[n]}; \epsilon)$ converges monotonically to $g_{p}({\bf{v}}^{\star}; \epsilon)$ for some KKT point ${\bf{v}}^{\star}$.
\end{enumerate}
\begin{IEEEproof}
Please refer to Appendix {\ref{profconvergence}} for details.
\end{IEEEproof} 
\end{theorem}
Compared with the algorithm for the smoothed-$\ell_p$ minimization in \cite{Yuanming_JSAC2015}, the main novelty of the proposed iterative reweighted-$\ell_2$ algorithm is revealing the explicit structures of the solutions at each iteration in Algorithm {\ref{irm1}}. Instead of using the interior-point algorithm to solve the convex subproblems at each iteration \cite{boyd2004convex}, Algorithm {\ref{irm1}} with closed-form solutions helps reduce the computational cost. {\RYS{The proposed iterative reweighted-$\ell_2$ algorithm can
only guarantee to converge to a KKT point, which may be a local minimum or
the other stationary point (e.g., a saddle point and local maximum).}}  

\begin{remark}
{\rev{The main contributions of the developed Algorithm {\ref{irm1}} include finding the closed-form solutions in the iterations, in comparison with the work \cite{Yuanming_JSAC2015} using the interior-point algorithm, and proving  the convergence of the low-complexity closed-form iterative algorithm for the non-convex group sparse inducing problem $\mathscr{P}_{\textrm{GSBF}}$ (\ref{slp}), instead of the convex coordinated beamforming problem \cite{WeiYu_WC10}.}} 
\end{remark}

Unfortunately,
computing the RRH ordering criterion $\theta_l$'s in (\ref{rrhorder}) requires to run Algorithm 1 for
each channel realization, which brings a heavy computation burden. In next
section,  we will find deterministic approximations for $\theta_l$'s
to determine the RRH ordering only based on statistical CSI, which changes
much more slowly than instantaneous channel states, and thus requires less frequent update.

\section{Group Sparse Beamforming via Large System Analysis}
\label{lsa}
In this section, we present the large system analysis for the iterative reweighted-$\ell_2$ algorithm in Algorithm {\ref{irm1}}, thereby enabling RRH ordering only based on statistical CSI. In this way, the ordering criterion will change only when the long-term channel attenuation is updated, and thus it can further reduce the computational complexity of group sparse beamforming. The main novelty of this section is providing the closed-forms for the  asymptotic analysis of the optimal Lagrangian multipliers based on the recent results in \cite{Couillet_2014large}, thereby providing explicit expressions for the asymptotic RRH ordering results. 

\subsection{Deterministic Equivalent of Optimal Parameters}
In this subsection, we provide asymptotic analysis for the optimal beamforming parameters $p_k$'s (\ref{dualpower}) and $\lambda_k$'s (\ref{dualpara}) when $N\rightarrow \infty$. In a Cloud-RAN with distributed RRHs, the channel  can be modeled as 
${\bf{h}}_k={\bs{\Theta}}_k^{1/2}{\bf{g}}_k, \forall k$,
where ${\bf{g}}_k\sim\mathcal{CN}({\bf{0}}, {\bf{I}}_{NL})$ is the small-scale fading and ${\bf{\Theta}}_k={\rm{diag}}\{d_{k1},\dots,
d_{kL}\}\otimes {\bf{I}}_N$ with $d_{kl}$ as the path-loss from RRH $l$
to MU $k$. With this channel model, we have the following result for the deterministic equivalent of the $\bs{\lambda}$-parameter.
\begin{lemma}[Asymptotic Results for $\bm{\lambda}$-Parameter]
\label{lemlam} 
Assume \[
0<\liminf\nolimits_{N\rightarrow
\infty} K/N\le 
\limsup_{N\rightarrow
\infty} K/N <\infty.
\] Let $\{d_{kl}\}$ and $\{\gamma_k\}$ satisfy $\limsup_{N}\max_{k,l}\{d_{kl}\}<\infty$
and $\lim\sup_N \max_k \gamma_k<\infty$, respectively.   We have $\max_{1\le k\le K}|\lambda_k-\lambda_k^{\circ}|\stackrel{N\rightarrow\infty}{\longrightarrow}
0$ almost surely, where $\lambda_k^{\circ}$ is given by
\begin{eqnarray}
\label{asymdual}
{\lambda}_k^{\circ}=\gamma_k\left({{1\over{L}}\sum_{l=1}^{L}d_{kl}
\eta_{l}}\right)^{-1}.
\end{eqnarray}
Here, $\eta_l$ is the unique solution of the following set of equations:
\begin{eqnarray}
\eta_l=\left({1\over{NL}}\sum_{i=1}^{K}{d_{il}\over{{1\over{L}}\sum_{j=1}^{L}d_{ij}\eta_j}}{\gamma_i\over{1+\gamma_i}}+\omega_l^{[n]}\right)^{-1}.
\end{eqnarray}
\begin{IEEEproof}
Please refer to Appendix {\ref{proflambda}} for details. 
\end{IEEEproof}
\end{lemma}

Based on the above results, we further have the following asymptotic result for the optimal powers $p_k$'s.
\begin{lemma}[Asymptotic Results for Optimal Powers]
\label{detpower}
Let ${\bf{\Delta}}\in\mathbb{C}^{K\times K}$ be such that
\begin{eqnarray}
[{\bf{\Delta}}]_{k,i}:={1\over{NL}}{\gamma_i\over{(1+\gamma_i)^2}}{\psi_{ik}'\over{\psi_i^{\circ2}}}.
\end{eqnarray}
If and only if $\limsup_K\|{\bf{\Delta}}\|_2<1$, then $\max_k|p_k-p_k^{\circ}|\stackrel{N\rightarrow\infty}{\longrightarrow}
0$ almost surely, where $p_k^{\circ}$ is given by
\begin{eqnarray}
\label{sympower}
p_k^{\circ}=\gamma_k{\psi_k'\over{\psi_k^{\circ2}}}\left({\tau_k\over{(1+\gamma_k)^2}}+\sigma_k^2\right).
\end{eqnarray}
Here $\psi_k^{\circ}$, $\psi_k'$ and $\psi_{ik}'$ are given as follows:
\begin{eqnarray}
\psi_k^{\circ}={1\over{L}}\sum_{l=1}^{L}d_{kl}\eta_l,
\end{eqnarray}
and
\begin{eqnarray}
\!\!\!\!\!\!\!\!\!\!\psi_{k}'={1\over{L}}\sum_{l=1}^Ld_{kl}\eta_l^2+{1\over{NL}}\sum_{j=1}^{K}{{{\lambda}_j^{\circ 2}\psi_{j}'}\over{(1+\gamma_j)^2}}{1\over{L}}\sum_{l=1}^{L}d_{il}d_{jl}\eta_l^2,
\end{eqnarray}
and
\begin{eqnarray}
\!\!\!\!\!\!\!\!\!\!\psi_{ik}'={1\over{L}}\sum_{l=1}^Ld_{il}d_{kl}\eta_l^2+{1\over{NL}}\sum_{j=1}^{K}{{{\lambda}_j^{\circ2}\psi_{jk}'}\over{(1+\gamma_j)^2}}{1\over{L}}\sum_{l=1}^{L}d_{il}d_{jl}\eta_l^2,
\end{eqnarray}
respectively; and ${\bs{\tau}}=[\tau_1, \dots, \tau_{K}]^T$
is given  as 
\begin{eqnarray}
{\bs{\tau}}=\sigma^2 \left({\bf{I}}_{K}-{\bf{\Delta}}\right)^{-1}{\bs{\delta}},
\end{eqnarray}
where ${\bs{\delta}}=[\delta_1,\dots, \delta_{K}]^T$ with
\begin{eqnarray}
\delta_k={1\over{NL}}\sum_{i=1}^{K}\gamma_i{\psi_{ik}'\over{\psi_i^{\circ2}}}.
\end{eqnarray} 
\begin{IEEEproof}
Please refer to Appendix {\ref{prodetpower}} for details. 
\end{IEEEproof}
\end{lemma}

\subsection{Statistical CSI Based Group Sparse Beamforming Algorithm}
Based on the deterministic equivalents of the optimal beamforming parameters $\lambda_k$'s and $p_k$'s in Lemma {\ref{lemlam}} and Lemma {\ref{detpower}}, respectively, we have the following theorem on the  squared $\ell_2$-norm of the solution to problem $\mathscr{P}_{\textrm{GSBF}}^{[n]}$, i.e.,
$\|\tilde{\bf{v}}_1^{[n]}\|_2^2, \dots, \|\tilde{\bf{v}}_L^{[n]}\|_2^2$.

\begin{theorem}[Asymptotic Results for the Beamformers] 
\label{detspa}
At the $n$-th iteration, for the  squared $\ell_2$-norm
of the solution to problem $\mathscr{P}_{\textrm{GSBF}}^{[n]}$,
$\|\tilde{\bf{v}}_1^{[n]}\|_2^2, \dots, \|\tilde{\bf{v}}_L^{[n]}\|_2^2$, we have $\max_{l}\left|\|\tilde{\bf{v}}_l^{[n]}\|_2^2-\bar\chi_l\right|\stackrel{N\rightarrow\infty}{\longrightarrow} 0$ almost surely with
\begin{eqnarray}
\label{detbeam}
\bar{\chi}_l={1\over{NL}}\sum_{k=1}^{K}{p}_k^{\circ}{\psi_{kl}\over{\psi_k'}},
\end{eqnarray}
where 
$\psi_{kl}={1\over{NL}}d_{kl}\eta_l^2+{1\over{NL}}\sum_{j=1}^{K}{{\lambda}_j^{\circ2}\psi_{jl}\over{(1+\gamma_j)^2}}{1\over{L}}\sum_{l=1}^Ld_{il}d_{jl}\eta_l^2$.
\begin{IEEEproof}
Please refer to Appendix \ref{prodetspa} for details. 
\end{IEEEproof}

\end{theorem}
 
 We thus have the following deterministic equivalent of the weights $\omega_l^{[n]}$'s (\ref{upweight}):
\begin{eqnarray}
\label{upweightdet}
\bar{\omega}_l^{[n]}= {{p\nu_l}\over{2}}\left[\bar{\chi}_l+\epsilon^2\right]^{{p\over{2}}-1},
\forall l=1,\dots, L.
\end{eqnarray}
Note that the asymptotic results of the  powers $p_k^{\circ}$'s (\ref{sympower})
and  Lagrangian multipliers $\lambda_k^{\circ}$'s (\ref{asymdual}) should
depend on the weights $\bar \omega_l^{[n]}$'s (\ref{upweightdet}) at each iteration. Based on (\ref{rrhorder}), we have the following asymptotic result for the RRH ordering criterion: 
\begin{eqnarray}
\label{rrhorderdet}
\bar\theta_l=\kappa_l\bar\chi_l^{\star},
\end{eqnarray} 
where  $\bar\chi_l^{\star}$ is the deterministic equivalent of the  squared $\ell_2$-norm
of the solution to problem $\mathscr{P}_{\textrm{GSBF}}$ using the iterative reweighted-$\ell_2$ algorithm. Therefore, the RRH with a smaller $\bar\theta_l$ will have a higher priority to be switched off. This ordering criterion will  change only when the long-term channel attenuation is updated. Note that global instantaneous CSI is still needed in stage II to find the active RRHs, i.e., solving a sequence of convex feasibility problems $\mathscr{F}(\mathcal{A}^{[i]})$ (\ref{feasiblecheck}).

The statistical CSI based iterative
reweighted-$\ell_2$ algorithm for group sparse beamforming is presented
in Algorithm {\ref{irm1sta}}. 
\begin{algorithm}
\label{irm1sta}
\caption{{Statistical CSI based Iterative Reweighted-$\ell_2$ Algorithm
for Problem $\mathscr{P}_{\textrm{GSBF}}$}}
{\textbf{input}}: Initialize ${\bs{\omega}}^{[0]}=(1, \dots, 1)$;
$I$ (the maximum number of iterations)\\
 Repeat
 
~~1) Compute deterministic equivalent of the squared $\ell_2$-norm of the solution to  $\mathscr{P}_{\textrm{GSBF}}^{[n]}$,
$\|\tilde{\bf{v}}_1^{[n]}\|_2^2, \dots, \|\tilde{\bf{v}}_L^{[n]}\|_2^2$,
using $\bar\chi_l^{[n]}$ (\ref{detbeam}).

~~2) Update the weights $\bar{\omega}_l^{[n+1]}$ using (\ref{upweightdet}). 

Until convergence or attain the  maximum iterations and return {\textbf{output}}.\\
 {\textbf{output}}: Asymptotic result of the RRH ordering criterion $\bar{\bm{\theta}}$ (\ref{rrhorderdet}).
\end{algorithm}
 
\begin{remark}
{\rev{Although stage 2 and stage 3 still require  instantaneous CSI, the RRH ordering criterion computed by Algorithm {\ref{irm1sta}} in stage 1 is only based on statistical CSI and thus will be updated only when the long-term channel propagation is changed. Therefore, Algorithm {\ref{irm1sta}} serves the purpose of further reducing the computation complexity of Algorithm {\ref{irm1}} for RRH ordering. A promising future research direction is to select RRHs in stage 2 only based on statistical CSI. However, the main challenge is the infeasibility issue if only statistical CSI is available, as discussed in Remark {\ref{reminfea}}. }}
\end{remark}

\section{Simulation Results}
\label{simres}
In this section, we will simulate the proposed algorithms for network power minimization for Cloud-RANs. In all the realizations, we only account the channel realizations making the original problem $\mathscr{P}$ feasible. If problem $\mathscr{P}$ is infeasible, further works on user admission are required \cite{Yuanming_JSAC2015}.  We set $p=1$, $\epsilon=10^{-3}$ and ${\bs{\omega}}^{[0]}=(1,\dots, 1)$ for all the simulations.  {\RYS{The proposed iterative reweighted-$\ell_2$ algorithms (Algorithm 1 and
Algorithm 2) will terminate if either the number of iterations exceeds 30
or the difference between the objective values of consecutive iterations
is less than $10^{-3}$.}}  

\subsection{Convergence of the Iteratively Reweighted-$\ell_2$ Algorithm}
\begin{figure}[t]
  \centering
  \includegraphics[width=0.9\columnwidth]{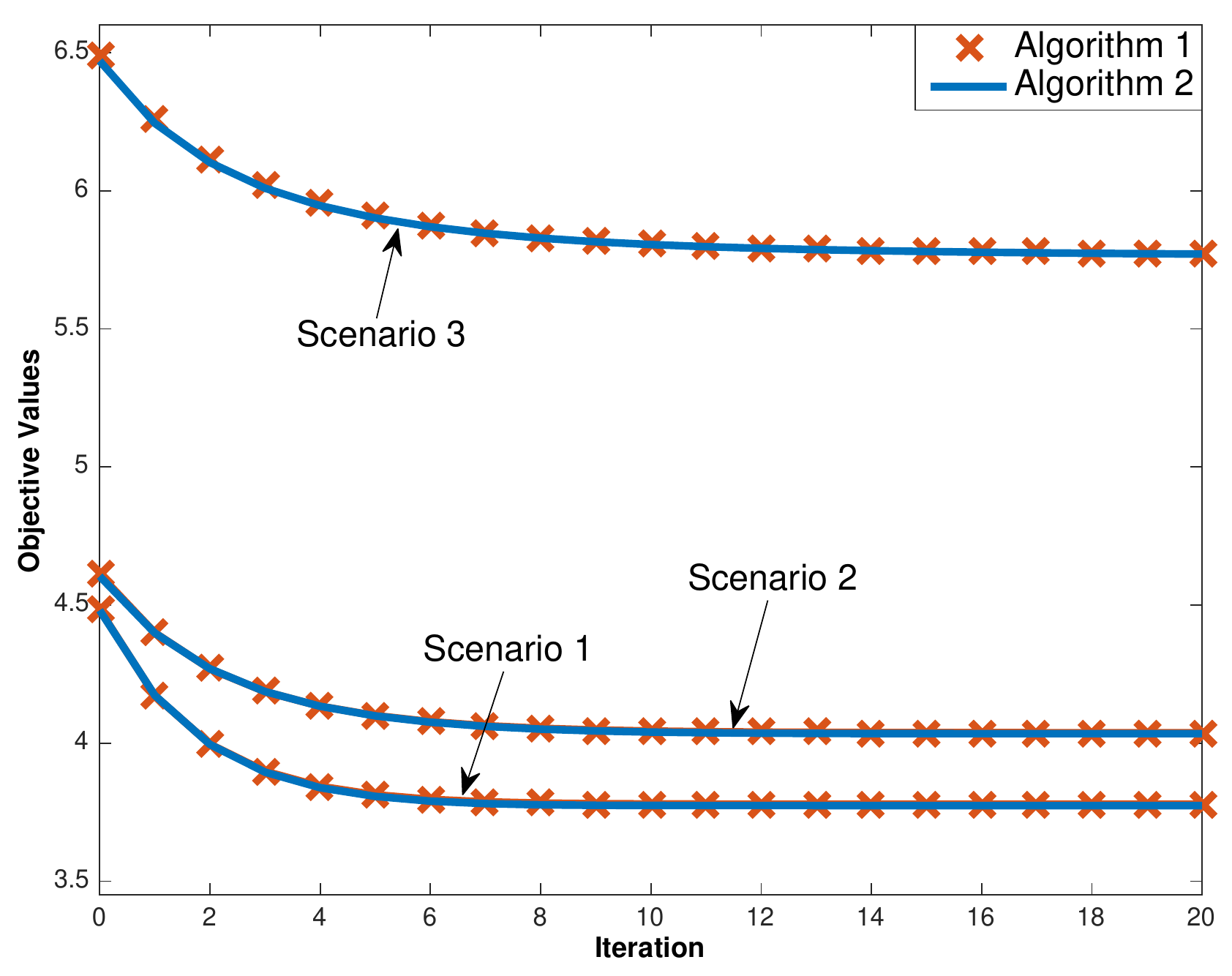}
 \caption{Convergence of the iterative reweighted-$\ell_2$ algorithm.}
 \label{cong}
\end{figure}
Consider a network with $L=5$ 30-antenna RRHs and 5 single antenna MUs uniformly and independently distributed in the square region $[-2000,2000]\times [-2000,2000]$ meters. Fig. {\ref{cong}} shows the convergence of the iterative reweighted-$\ell_2$ algorithm {\RYS{in different scenarios with different realizations of RRH and MU positions}}. {\RYS{For each scenario,}} the numerical results are obtained by averaging over 100 small-scale fading realizations. The large-scale fading (i.e., statistical CSI) is randomly generated and fixed during simulations. In the two curves {\RYS{of each scenario}}, Algorithm 1 and Algorithm 2 are applied for Stage I of group sparse beamforming, respectively. It demonstrates that the large system analysis based Algorithm \ref{irm1sta} provides accurate approximation even in a small system. Fast convergence is observed in the simulated setting.   

\subsection{Network Power Minimization}
 \begin{table}[!t]
\renewcommand{\arraystretch}{1.3}
\caption{Network Power Consumption using Algorithm 1 and Algorithm 2 with different parameters $p$}
\label{parameterp}
\centering
\begin{tabular}{l|c|c|c|c|c}
\hline
\tabincell{c}{{Target SINR [dB]}} & 0 & 3 &
6 & 9 & 12 \\
\hline
GSBF [7] & 27.671 & 28.528 & 29.924 & 32.759 & 43.855  \\
\hline
\tabincell{c}{Alg. 1 with $p=1$} & 23.716 & 24.374 & 26.011
&30.057 & 43.810 \\
\hline
\tabincell{c}{Alg. 1 with $p=0.5$} & 23.789 & 24.463 & 26.092 &
30.500 & 43.783  \\
\hline
\tabincell{c}{Alg. 2 with $p=1$} & 23.907 & 24.695 & 26.326 & 30.214 & 43.827  \\
\hline
\tabincell{c}{Alg. 2 with $p=0.5$} & 23.905 & 24.689 & 26.315 & 30.205 & 43.821\\
\hline
\end{tabular}
\label{rrh_num}
\end{table}

\begin{figure}[t]
  \centering
  \includegraphics[width=0.9\columnwidth]{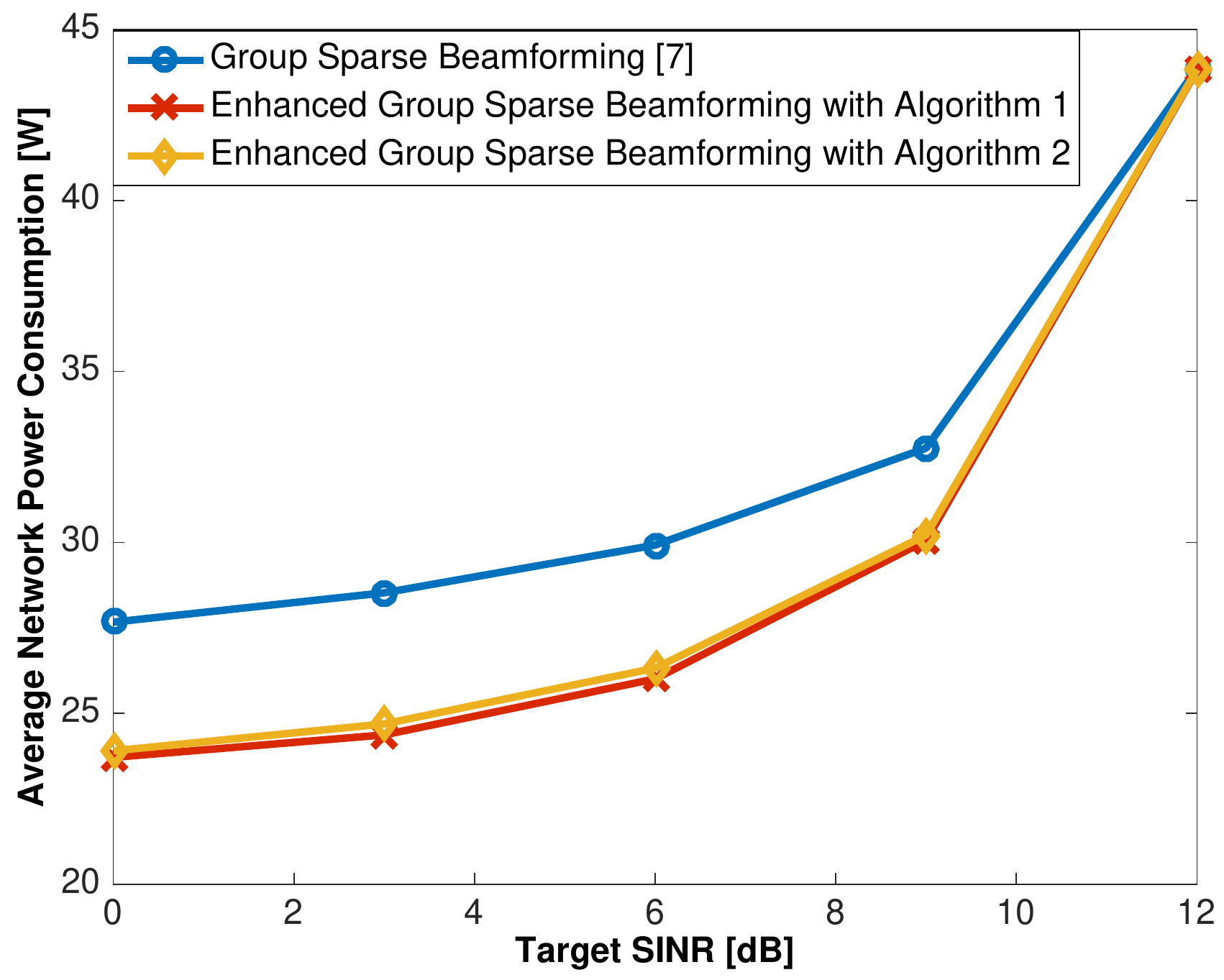}
 \caption{Average network power consumption versus target SINR with different algorithms.}
 \label{netp}
\end{figure}  
Consider a network with $L=5$ 10-antenna RRHs and 6 single antenna MUs uniformly
and independently distributed in the square region $[-1000,1000]\times [-1000,1000]$
meters. The relative transport link power consumption are set to be $P_l^c=(5.6+2l)
W, l=1,\dots, L$. We average over 200 small-scale channel realizations. Fig. {\ref{netp}} demonstrates the average network power consumption using different algorithms. This figure shows that the proposed iterative reweighted-$\ell_2$ algorithm outperforms the weighted mixed $\ell_1/\ell_2$-norm algorithm in \cite{Yuanming_TWC2014}. Furthermore, it illustrates that the large system analysis provides accurate approximations for the iterative reweighted-$\ell_2$ algorithm in finite systems with reduced computation overhead. {\RYS{In Table {\ref{parameterp}}, for the simulated scenarios, we see that the iterative reweighted-$\ell_2$ algorithms achieve similar performance with different values of parameter $p$ and the large system analysis provides accurate approximations. Although the simulated results demonstrate that the performance is robust to different values of   parameter $p$, it is very interesting to theoretically identify the typical scenarios, where smaller values of parameter $p$ will yield much lower network power consumption.}}

\section{Conclusions and Discussions}
\label{condis}
In this paper, we developed an enhanced three-stage group sparse beamforming framework  with reduced computation overhead in Cloud-RANs. In particular, closed-form solutions for the group sparse optimization problem were obtained by developing the iterative reweighted-$\ell_2$ algorithm based on the MM algorithm and Lagrangian duality theory. This is the first effort to reduce the computation cost for RRH ordering in the first stage of group sparse beamforming. Based on the developed structured iterative algorithm, we further provided a large system analysis for the optimal Lagrangian multipliers at each iteration via random matrix theory, thereby computing the RRH ordering criterion only based on the statistical CSI. This is the second effort to enable computation scalability for RRH ordering in the first stage.   

Several future directions of interest are listed as follows:
\begin{itemize}
\item Although computing the RRH ordering criterion in the first stage with Algorithm \ref{irm1sta} only needs  statistical CSI, the second stage still requires global instantaneous CSI to find the active RRHs by solving a sequence of convex feasibility problems. It is thus particularly interesting to investigate efficient algorithms to select the active RRHs only based on the statistical CSI. 

\item It is desirable to establish the optimality of the iterative reweighted-$\ell_2$ algorithm for the network power minimization problem $\mathscr{P}$. However, considering the complicated problem structures of problem $\mathscr{P}$, this becomes challenging. {\RYS{It is also interesting to apply the developed statistical CSI based iterative reweighted-$\ell_2$ algorithm for more complicated network optimization problems, e.g., the wireless caching problem \cite{Jun_caching2014, Tao_TWC16}, the computation offloading problem \cite{Yuanming_Compoffloading}, and beamforming problems with CSI uncertainty \cite{Yuanming_SDP2014}.}}   
\end{itemize}
       
\appendices

\section{Proof of Theorem {\ref{tconvergence}}}
\label{profconvergence}
1) {\RYS{Observing that the phases of ${\bf{v}}_k$ will not change
the objective function and constraints of problem $\mathscr{P}_{\textrm{GSBF}}$,
the QoS constraints thus can be equivalently transformed to the  second-order
cone constraints: 
$\mathcal{C}=\left\{{\bf{v}}:\sqrt{\sum\nolimits_{i\ne k}|{\bf{h}}_{k}^{\sf{H}}{\bf{v}}_i|^2+\sigma_k^2}\le \mathfrak{R}({\bf{h}}_k^{\sf{H}}{\bf{v}}_k)/\gamma_k, \forall k\right\}$,
which are convex cones.}} The KKT points of problem $\mathscr{P}_{\textrm{GSBF}}$ should satisfy:
\begin{eqnarray}
\label{kktd}
0\in\nabla_{\bf{v}}g_{p}({\bf{v}}; \epsilon)+\mathcal{N}_{\mathcal{C}}(\bf{v}),
\end{eqnarray}
where $\mathcal{N}_{\mathcal{C}}(\bf{v})$ is the normal cone of the second-order cone at point $\bf{v}$ \cite{Rockafellar1997convex}. We shall show that any convergent subsequence $\{{\bf{v}}^{[n_k]}\}_{k=1}^{\infty}$ of $\{{\bf{v}}^{[n]}\}_{n=1}^{\infty}$ satisfies (\ref{kktd}). Specifically, let ${\bf{v}}^{[n_k]}\rightarrow\bar{\bf{v}}$ be one such convergent subsequence with 
\begin{eqnarray}
\label{limseq}
\lim_{k\rightarrow\infty}{\bf{v}}^{[n_k+1]}=\lim_{k\rightarrow\infty}{\bf{v}}^{[n_k]}=\bar{\bf{v}}.
\end{eqnarray}
Based on the strong duality result for problem $\mathscr{P}_{\textrm{GSBF}}^{[n]}$, the following KKT condition holds at ${\bf{v}}^{[n_k+1]}$:
\begin{eqnarray}
\label{kkt22}
0\in\nabla_{\bf{v}}Q({\bf{v}}^{[n_k+1]}; {\bm{\omega}}^{[n_k]})+\mathcal{N}_{\mathcal{C}}({\bf{v}}^{[n_k+1]}).
\end{eqnarray}
Based on (\ref{limseq}) and (\ref{upweight}), we have
$\lim_{k\rightarrow\infty}\nabla_{\bf{v}}Q({\bf{v}}^{[n_k+1]}; {\bm{\omega}}^{[n_k]})=\nabla_{\bf{v}}g_{p}(\bar{\bf{v}}; \epsilon)$.
Furthermore, based on \cite[Proposition 6.6]{Rockafellar1997convex} and (\ref{limseq}), we have 
$\limsup_{{\bf{v}}^{[n_k+1]}\rightarrow\bar{\bf{v}}}\mathcal{N}_{\mathcal{C}}({\bf{v}}^{[n_k+1]})=\mathcal{N}_{\mathcal{C}}(\bar{\bf{v}})$.
Therefore, by taking $k\rightarrow\infty$ in (\ref{kkt22}), we have
$0\in \nabla_{\bf{v}}Q(\bar{\bf{v}}; \bar{\bm{\omega}})+\mathcal{N}_{\mathcal{C}}(\bar{\bf{v}})$,
which indicates that $\bar{\bf{v}}$ is a KKT point of problem $\mathscr{P}_{\textrm{GSBF}}$. We thus complete the proof.

2) We first conclude the following fact:
\begin{eqnarray}
\label{ub1}
&&g_p({\bf{v}}^{[n+1]};\epsilon)\nonumber\\
&=&Q({\bf{v}}^{[n+1]};{\bs{\omega}}^{[n]})+g_p({\bf{v}}^{[n+1]};\epsilon)-Q({\bf{v}}^{[n+1]};
{\bs{\omega}}^{[n]})\nonumber\\
&\le& Q({\bf{v}}^{[n+1]};{\bs{\omega}}^{[n]})+g_p({\bf{v}}^{[n]};\epsilon)-Q({\bf{v}}^{[n]};
{\bs{\omega}}^{[n]})\nonumber\\
&\le& Q({\bf{v}}^{[n]}; {\bs{\omega}}^{[n]})+g_p({\bf{v}}^{[n]};\epsilon)-Q({\bf{v}}^{[n]};
{\bs{\omega}}^{[n]})\nonumber\\
&=&g_p({\bf{v}}^{[n]};\epsilon),
\end{eqnarray}
where the first inequality is based on the fact that function $(g_{p}({\bf{v}};\epsilon)-Q({\bf{v}};
{\bs{\omega}}^{[n]}))$ attains its maximum at ${\bf{v}}={\bf{v}}^{[n]}$ \cite[Proposition 1]{Yuanming_JSAC2015},
and the second inequality follows from (\ref{kkt22}). Furthermore, as $g_p({\bf{v}}; \epsilon)$ is continuous and $\mathcal{C}$ is compact, the limit of the sequence $g_p({\bf{v}}^{[n]};
\epsilon)$ is finite. Based on the convergence results in 1), we thus complete the proof.

\section{Proof of Lemma {\ref{lemlam}}}
\label{proflambda}
{\rev{The proof technique here is mainly based on the work  \cite{Couillet_TWC16}. However, we have to modify their proof as we have different weighting matrices ${\bf{Q}}^{[n]}$ in vectors $\bar{\bf{v}}_k$ (\ref{bfs1}) at the $n$-th iteration and such modification is non-trivial.}} Based on the following matrix inversion lemma \cite{horn2012matrix}, 
${\bf{x}}^{\sf{H}}({\bf{U}}+c{\bf{x}}{\bf{x}}^{\sf{H}})^{-1}={{\bf{x}}^{\sf{H}}{\bf{U}}^{-1}\over{1+c{\bf{x}}^{\sf{H}}{\bf{U}}^{-1}{\bf{x}}}}$,
(\ref{dualpara}) can be rewritten as
\begin{eqnarray}
\label{lamre}
{\gamma_k\over{\lambda_k}}={1\over{LN}}{\bf{h}}_k^{\sf{H}}\left({1\over{LN}}\sum_{
i\ne k}{{\lambda_i}}{\bf{h}}_i{\bf{h}}_i^{\sf{H}}+{\bf{Q}}\right)^{-1}{\bf{h}}_k,
\end{eqnarray}
which can be further rewritten as
\begin{eqnarray}
\label{dualparare}
\gamma_k\rho_k={{\lambda}_k^{\circ}\over{LN}}{\bf{h}}_k^{\sf{H}}\left({1\over{LN}}\sum_{
i\ne
k}{{{\lambda}_i^{\circ}}\over{\rho_i}}{\bf{h}}_i{\bf{h}}_i^{\sf{H}}+{\bf{Q}}\right)^{-1}{\bf{h}}_k,
\end{eqnarray}
where $\rho_k={{\lambda}_k^{\circ}/{\lambda_k}}$.

Assume that $0\le \rho_1\le \rho_2\le\cdots\le \rho_K$. From (\ref{dualparare}),
replacing $\rho_i$ with $\rho_K$ and using monotonicity
arguments, we have
\begin{eqnarray}
\gamma_K\rho_K\le {{\lambda}_K^{\circ}\over{LN}}{\bf{h}}_K^{\sf{H}}\left({1\over{LN}}\sum_{
i\ne K}{{{\lambda}_i^{\circ}}\over{\rho_K}}{\bf{h}}_i{\bf{h}}_i^{\sf{H}}+{\bf{Q}}\right)^{-1}{\bf{h}}_K,
\end{eqnarray}
or, equivalently
\begin{eqnarray}
\label{rhoine}
\gamma_{K}\le {{\lambda}_K^{\circ}\over{LN}}{\bf{h}}_K^{\sf{H}}\left({1\over{LN}}\sum_{
i\ne
K}{{{\lambda}_i^{\circ}}}{\bf{h}}_i{\bf{h}}_i^{\sf{H}}+\rho_K{\bf{Q}}\right)^{-1}{\bf{h}}_K.
\end{eqnarray}
Assume now that $\rho_K$
is infinitely
often larger than $(1+\ell)$ with $\ell$ being some positive value \cite{Couillet_2014large}.
Let us restrict ourselves to such a subsequence. From (\ref{rhoine}), using
monotonicity arguments we obtain
\begin{eqnarray}
\label{rhoine1}
\gamma_{K}\le {{\lambda}_K^{\circ}\over{LN}}{\bf{h}}_K^{\sf{H}}\left({1\over{LN}}\sum_{
i\ne
K}{{{\lambda}_i^{\circ}}}{\bf{h}}_i{\bf{h}}_i^{\sf{H}}+(1+\ell){\bf{Q}}\right)^{-1}\!\!\!\!\!\!\!{\bf{h}}_K.
\end{eqnarray}

Define
\begin{eqnarray}
\!\!\!\!\!m_k(\ell)={1\over{LN}}{\bf{h}}_k^{\sf{H}}\left({1\over{LN}}\sum_{
i\ne
k}{{{\lambda}_i^{\circ}}}{\bf{h}}_i{\bf{h}}_i^{\sf{H}}+(1+\ell){\bf{Q}}\right)^{-1}\!\!\!\!\!\!\!{\bf{h}}_k.
\end{eqnarray}
Now we investigate the deterministic equivalents for $m_k(\ell)$. We first
introduce the following important lemma:
\begin{lemma}\cite[Lemma 14.2]{Couillet_2011random}
\label{quatrac}
Let ${\bf{A}}_1, {\bf{A}}_2,\dots,$ with ${\bf{A}}_N\in\mathbb{C}^{N\times
N}$ be a series of random matrices generated by the probability space $(\Omega,
\mathcal{F}, P)$ such that, for $\omega\in A\subset \Omega$, with $P(A)=1,
\|{\bf{A}}_N(\omega)\|< K(\omega)<\infty$, uniformly on $N$. Let ${\bf{x}}_1,
{\bf{x}}_2,\dots,$ with ${\bf{x}}_N\in\mathbb{C}^N$, be random vectors of
i.i.d. entries with zero mean, variance $1/N$, and the eighth-order moment of
order $O(1/N^4)$, independent of ${\bf{A}}_N$. Then
\begin{eqnarray}
{\bf{x}}_N^{\sf{H}}{\bf{A}}_N{\bf{x}}_N-{1\over{N}}{\rm{Tr}}({\bf{A}}_N)\stackrel{N\rightarrow\infty}{\longrightarrow}
0,
\end{eqnarray}
almost surely. 
\end{lemma}
Therefore, based on Lemma {\ref{quatrac}}, we have
\begin{eqnarray}
&&m_K(\ell)-{1\over{LN}}{\rm{Tr}}{\bf{\Theta}}_K\left({1\over{LN}}\sum_{
i\ne K}{{{\lambda}_i^{\circ}}}{\bf{h}}_i{\bf{h}}_i^{\sf{H}}+(1+\ell){\bf{Q}}\right)^{-1}\nonumber\\
&&\stackrel{N\rightarrow\infty}{\longrightarrow} 0.
\end{eqnarray}

We further introduce the following important lemma to derive the deterministic
equivalent for the $m_k(l)$.
\begin{lemma} \cite[Theorem 1]{Debbah_TIT2012}
\label{detthe}
Let ${\bf{B}}_N={\bf{X}}_N^{\sf{H}}{\bf{X}}_N+{\bf{S}}_N$ with ${\bf{S}}_N\in\mathbb{C}^{N\times
N}$ Hermitian nonnegative definite and ${\bf{X}}_N\in\mathbb{C}^{n\times
N}$ random. The $i$-th column ${\bf{x}}_i$ of ${\bf{X}}_N^{\sf{H}}$ is ${\bf{x}}_i={\bf{\Psi}}_i{\bf{y}}_i$,
where the entries of ${\bf{y}}_i\in\mathbb{C}^{r_i}$ are i.i.d. of zero mean,
variance $1/N$ and have eighth-order moment of order $O(1/N^4)$. The matrices
${\bf{\Psi}}_i\in\mathbb{C}^{N\times r_i}$ are deterministic. Furthermore,
let ${\bf{\Theta}}_i={\bf{\Psi}}_i{\bf{\Psi}}_i^{\sf{H}}\in\mathbb{C}^{N\times
N}$ and define ${\bf{Q}}_N\in\mathbb{C}^{N\times N}$ deterministic. Assume
$\limsup_{N\rightarrow\infty}\sup_{1\le i\le n}\|{\bf{\Theta}}_i\|<\infty$
and let ${\bf{Q}}_N$ have uniformly bounded spectral norm (with respect to
$N$). Define
\begin{eqnarray}
m_{{\bf{B}}_N, {\bf{Q}}_N}(z):={1\over{N}}{\rm{Tr}}{\bf{Q}}_N({\bf{B}}_N-z{\bf{I}}_N)^{-1}.
\end{eqnarray}
Then, for $z\in\mathbb{C}\backslash\mathbb{R}^{+}$, as $n, N$ grow large
with ratios $\beta_{N,i}:=N/r_i$ and $\beta_N:=N/n$ such that $0<\liminf_N\beta_N\le
\lim\sup_N\beta_N<\infty$ and $0<\liminf_N\beta_{N,i}\le\limsup_{N}\beta_{N,i}<\infty$,
we have that
\begin{eqnarray}
m_{{\bf{B}}_N, {\bf{Q}}_N}(z)-m_{{\bf{B}}_N, {\bf{Q}}_N}^{\circ}(z)\stackrel{N\rightarrow\infty}{\longrightarrow}
0
\end{eqnarray}
almost surely, with $m_{{\bf{B}}_N, {\bf{Q}}_N}^{\circ}(z)$ given by
\begin{eqnarray}
&&m_{{\bf{B}}_N, {\bf{Q}}_N}^{\circ}(z)\nonumber\\
&&={1\over{N}}{\rm{Tr}}{\bf{Q}}_N\left({1\over{N}}\sum_{j=1}^{n}{{{\bf{\Theta}}_j}\over{1+e_{N,
j(z)}}}+{\bf{S}}_N-z{\bf{I}}_N\right)^{-1},
\end{eqnarray}
where the functions $e_{N,1}(z),\dots, e_{N,n}(z)$ from the unique solution
of 
\begin{eqnarray}
\label{stj}
\!\!\!\!\!\!\!e_{N,i}(z)={1\over{N}} {\rm{Tr}}{{\bf{\Theta}}_{i}}\left({1\over{N}}\sum_{j=1}^{n}{{{\bf{\Theta}}_j}\over{1+e_{N,j}(z)}}+{\bf{S}}_N-z{\bf{I}}_N\right)^{-1}\!\!\!\!\!\!\!,
\end{eqnarray}
which is the Stieltjes transform of a nonnegative finite measure on $\mathbb{R}^{+}$.
Moreover, for $z<0$, the scalars $e_{N,1}(z), \dots, e_{N,n}(z)$ are unique
nonnegative solutions to (\ref{stj}).
\end{lemma}
Therefore, based on Lemma {\ref{detthe}}, we have
\begin{eqnarray}
\label{detres1}
\psi_K(\ell)-\psi_{K}^{\circ}(\ell)\stackrel{N\rightarrow\infty}{\longrightarrow}
0
\end{eqnarray}
almost surely with $\psi_K^{\circ}(\ell)$ being the unique positive solution
to
\begin{eqnarray}
\psi_K^{\circ}(\ell)={1\over{NL}}{\rm{Tr}}({\bf{\Theta}}_K{\bf{A}}(\ell))
\end{eqnarray} 
where 
\begin{eqnarray}
\label{al}
{\bf{A}}(\ell)=\left({1\over{NL}}\sum_{i=1}^{K}{{{\lambda}}_i^{\circ}{\bf{\Theta}}_i\over{1+{\lambda}}_i^{\circ}\psi_i^{\circ}(\ell)}+(1+\ell){\bf{Q}}\right)^{-1}.
\end{eqnarray}

From (\ref{detres1}), recalling (\ref{rhoine1}) yields 
\begin{eqnarray}
\lim_{K\rightarrow\infty}\inf
\psi_K^{\circ}(\ell)\ge{\gamma_K/{{\lambda}_K^{\circ}}}
\end{eqnarray}
Using the fact that $\psi_K^{\circ}(\ell)$ is a decreasing function of $\ell$, it can be
proved \cite{Couillet_2011random} that for any $\ell>0$, we have 
\begin{eqnarray}
\lim_{K\rightarrow\infty}\sup
\psi_K^{\circ}(\ell)< \gamma_K/{\lambda}_K^{\circ}.
\end{eqnarray}
However, this is against the former
condition and creates a contradiction on the initial hypothesis that $\rho_K
< 1+\ell$ infinitely often. Therefore, we must admit that $\rho_K\le 1+\ell$
for all large values of $K$. Reverting all inequalities and using similar
arguments yields $\rho_K\ge 1-\ell$ for all large values of $K$. 

We eventually obtain that $1-\ell \le \rho_{K} \le 1+\ell$ from which we
may write $\max|\rho_K-1|\le \ell$ for all large values of $K$. Taking a
countable sequence of $\ell$ going to zero yields $\max |\rho_K-1|\rightarrow
0$ from which using $\rho_K=\lambda_K^{\circ}/\lambda_K$ and assuming
$\lim_{K\rightarrow \infty} \sup \gamma_K <\infty$, we obtain
\begin{eqnarray}
\max |\lambda_K-{\lambda}_K^{\circ}|\stackrel{N\rightarrow\infty}{\longrightarrow} 0,
\end{eqnarray} 
almost surely. From (\ref{dualparare}), we have $\gamma_K/\lambda_K^{\circ}=\psi_K^{\circ}(0)=\psi_K^\circ$.
Therefore, we obtain 
$\lambda_K^{\circ}=\gamma_K/\psi_K^{\circ}$,
where $\psi_K^{\circ}$ is the solution of the following equation:
\begin{eqnarray}
\psi_K^{\circ}={1\over{NL}}{\rm{Tr}}{\bf{\Theta}}_K\left({1\over{NL}}\sum_{i=1}^{K}{{{\bf{\Theta}}_i}\over{\psi_K^{\circ}}}{{{\gamma}}_i\over{1+{\gamma}}_i}+{\bf{Q}}\right)^{-1}.
\end{eqnarray}
Following the same steps for $k=1,\dots, K$ yields the following desired result:
\begin{eqnarray}
\lambda_k^{\circ}=\gamma_k/\psi_k^{\circ}, \forall k.
\end{eqnarray}  

As ${\bf{\Theta}}_k={\rm{diag}}\{d_{k1},\dots,
d_{kL}\}\otimes {\bf{I}}_N$, we have 
\begin{eqnarray}
\label{psizero}
\psi_k^{\circ}={1\over{LN}}{\rm{Tr}}({\bf{\Theta}}_k{\bf{A}})={1\over{L}}\sum_{l=1}^{L}d_{kl}
\eta_{l},
\end{eqnarray}
where ${\bf{A}}={\bf{A}}(0)$ in (\ref{al}) and $\{\eta_l\}$ is the unique positive solution to the following set of
equations
\begin{eqnarray}
\eta_l=\left({1\over{NL}}\sum_{i=1}^{K}{d_{il}\over{1/L\sum_{j=1}^{L}d_{kj}\eta_j}}{\gamma_i\over{1+\gamma_i}}+\omega_l^{[n]}\right)^{-1}.
\end{eqnarray}
We thus complete the proof for the asymptotic results of the optimal Lagrange
multipliers in Lemma {\ref{lemlam}}.

\section{Proof of Lemma {\ref{detpower}}}
\label{prodetpower}
The SINR for MU $k$ (\ref{sinrexp}) can be rewritten as
\begin{eqnarray}
\label{sinrnew}
{\sf{sinr}}_{k}={{{p_k\over{NL}}{|{\bf{h}}_{k}^{\sf{H}}\tilde{\bf{v}}_{k}|^2\over{\|\tilde{\bf{v}}_k\|_2^2}}}\over{\sum_{i\ne
k}{{p_i}\over{NL}}{|{\bf{h}}_{k}^{\sf{H}}\tilde{\bf{v}}_{i}|^2\over{\|\tilde{\bf{v}}_i\|_2^2}}+\sigma_{k}^2}},
\end{eqnarray}
where $\tilde{\bf{v}}_k=\left(\sum_{i=1}^{K}{{{\lambda}_i^{\circ}}\over{LN}}{\bf{h}}_i{\bf{h}}_i^{\sf{H}}+{\bf{Q}}^{1/2}\right)^{-1}{\bf{h}}_k$.
The interference part of the denominator of ${\Gamma}_k$ in (\ref{sinrnew}) can be rewritten as
\begin{eqnarray}
\label{sinrde}
&&{1\over{NL}}\sum_{i\ne k} p_i{{|{\bf{h}}_k^{\sf{H}}\tilde{\bf{v}}_i|^2}\over{\|\tilde{\bf{v}}_i\|_2^2}}={1\over{NL}}\sum_{i\ne
k}p_i{\bf{h}}_k^{\sf{H}}\left({{\tilde{\bf{v}}_i\tilde{\bf{v}}_i^{\sf{H}}}\over{\|\tilde{\bf{v}}_i\|_2^2}}\right){\bf{h}}_k\nonumber\\
&=&{1\over{NL}}{\bf{h}}_k^{\sf{H}}{\bf{V}}\left({1\over{NL}}{\bf{H}}^{[k]}{\bf{P}}^{[k]}{\bf{H}}^{[k]\sf{H}}\right){\bf{V}}{\bf{h}}_k
\end{eqnarray}
with ${\bf{V}}=\left({1\over{NL}}\sum_{i=1}^K\lambda_i^{\circ}{\bf{h}}_i{\bf{h}}_i^{\sf{H}}+{\bf{Q}}\right)^{-1}$,
${\bf{H}}^{[k]}:=[{\bf{h}}_1,\dots, {\bf{h}}_{k-1}, {\bf{h}}_{k+1}, \dots,
{\bf{h}}_K]\in\mathbb{C}^{NL\times (K-1)}$ and
${\bf{P}}^{[k]}:={\rm{diag}}\left\{{p_1\over{{1\over{NL}}\|\tilde{\bf{v}}_1\|_2^2}},\dots,
{p_{k-1}\over{{1\over{NL}}\|\tilde{\bf{v}}_{k-1}\|_2^2}}, {p_{k+1}\over{{1\over{NL}}\|\tilde{\bf{v}}_{k+1}\|_2^2}},\dots,
{p_K\over{{1\over{NL}}\|\tilde{\bf{v}}_K\|_2^2}}\right\}$.

In order to eliminate the dependence between ${\bf{h}}_k$ and ${\bf{V}}$,
rewrite (\ref{sinrde}) as
\begin{eqnarray}
\label{sinrde1}
&&{1\over{NL}}{\bf{h}}_k^{\sf{H}}{\bf{V}}\left({1\over{NL}}{\bf{H}}^{[k]}{\bf{P}}^{[k]}{\bf{H}}^{[k]\sf{H}}\right){\bf{V}}{\bf{h}}_k\nonumber\\
&&={1\over{NL}}{\bf{h}}_k^{\sf{H}}{\bf{V}}^{[k]}\left({1\over{NL}}{\bf{H}}^{[k]}{\bf{P}}^{[k]}{\bf{H}}^{[k]\sf{H}}\right){\bf{V}}{\bf{h}}_k+\nonumber\\
&&{1\over{NL}}{\bf{h}}_k^{\sf{H}}\left({\bf{V}}-{\bf{V}}^{[k]}\right)\left({1\over{NL}}{\bf{H}}^{[k]}{\bf{P}}^{[k]}{\bf{H}}^{[k]\sf{H}}\right){\bf{V}}{\bf{h}}_k,
\end{eqnarray}
where ${\bf{V}}^{[k]}=\left({1\over{NL}}\sum_{i\ne k}{\lambda}_i^{\circ}{\bf{h}}_i{\bf{h}}_i^{\sf{H}}+{\bf{Q}}\right)^{-1}$.
Using the resolvent identity \cite{horn2012matrix} (i.e., ${\bf{U}}^{-1}-{\bf{V}}^{-1}=-{\bf{U}}^{-1}({\bf{U}}-{\bf{V}}){\bf{V}}^{-1}$
with ${\bf{U}}$ and ${\bf{V}}$ as two invertible complex matrices of size
$N\times N$), we have 
\begin{eqnarray}
{\bf{V}}-{\bf{V}}^{[k]}=-{\bf{V}}({\bf{V}}^{-1}-{\bf{V}}^{[k]-1}){\bf{V}}^{[k]}.
\end{eqnarray}
Then, observing that 
${\bf{V}}^{-1}-{\bf{V}}^{[k]-1}={{\lambda}_k^{\circ}\over{NL}}{\bf{h}}_k{\bf{h}}_k^{\sf{H}}$.
From (\ref{sinrde1}), one gets 
\begin{eqnarray}
\label{sinrde2}
&&{1\over{NL}}{\bf{h}}_k^{\sf{H}}{\bf{V}}\left({1\over{NL}}{\bf{H}}^{[k]}{\bf{P}}^{[k]}{\bf{H}}^{[k]\sf{H}}\right){\bf{V}}{\bf{h}}_k\nonumber\\
&=&{1\over{NL}}{\bf{h}}_k^{\sf{H}}{\bf{V}}^{[k]}\left({1\over{NL}}{\bf{H}}^{[k]}{\bf{P}}^{[k]}{\bf{H}}^{[k]\sf{H}}\right){\bf{V}}{\bf{h}}_k-\nonumber\\
&&{\bar{\lambda}_k\over{NL}}{\bf{h}}_k^{\sf{H}}{\bf{V}}{\bf{h}}_k\left[{1\over{NL}}{\bf{h}}_k^{\sf{H}}{\bf{V}}^{[k]}\left({1\over{NL}}{\bf{H}}^{[k]}{\bf{P}}^{[k]}{\bf{H}}^{[k]\sf{H}}\right){\bf{V}}{\bf{h}}_k\right].\nonumber\\
\end{eqnarray}

\begin{lemma}\cite[Lemma 7]{Debbah_TIT2012} 
\label{qualem}
Let ${\bf{U}}, {\bf{V}}, {\bf{\Theta}}\in\mathbb{C}^{N\times N}$ be of uniformly
bounded spectral norm with respect to $N$ and let $\bf{V}$ be invertible.
Further, define ${\bf{x}}:={\bf{\Theta}}^{1/2}{\bf{z}}$ and ${\bf{y}}:={\bf{\Theta}}^{1/2}{\bf{q}}$,
where ${\bf{z}}, {\bf{q}}\in\mathbb{C}^{N}$ have i.i.d. complex entries of
zero mean, variance $1/N$, and finite eighth-order moment and be mutually
independent as well as independent of ${\bf{U}}, {\bf{V}}$. Define $c_0,
c_1, c_2\in\mathbb{R}^+$ such that $c_0c_1-c_2^2\ge 0$ and let $\mu:={1\over{N}}{\rm{Tr}}{\bf{\Theta}}{\bf{V}}^{-1}$
and $\mu':={1\over{N}}{\rm{Tr}}({\bf{\Theta}}{\bf{U}}{\bf{V}}^{-1})$. Then,
we have
\begin{eqnarray}
&&{\bf{x}}^{\sf{H}}{\bf{U}}({\bf{V}}+c_0{\bf{x}}{\bf{x}}^{\sf{H}}+c_1{\bf{y}}{\bf{y}}^{\sf{H}}+c_2{\bf{x}}{\bf{y}}^{\sf{H}}+c_2{\bf{y}}{\bf{x}}^{\sf{H}})^{-1}{\bf{x}}-\nonumber\\
&&{{\mu'(1+c_1\mu)}\over{(c_0c_1-c_2^2)\mu^2+(c_0+c_1)\mu+1}}\stackrel{N\rightarrow\infty}{\longrightarrow}
0
\end{eqnarray}
almost surely. Furthermore,
\begin{eqnarray}
&&{\bf{x}}^{\sf{H}}{\bf{U}}({\bf{V}}+c_0{\bf{x}}{\bf{x}}^{\sf{H}}+c_1{\bf{y}}{\bf{y}}^{\sf{H}}+c_2{\bf{x}}{\bf{y}}^{\sf{H}}+c_2{\bf{y}}{\bf{x}}^{\sf{H}})^{-1}{\bf{y}}-\nonumber\\
&&{{c_2\mu\mu'}\over{(c_0c_1-c_2^2)\mu^2+(c_0+c_1)\mu+1}}\stackrel{N\rightarrow\infty}{\longrightarrow}
0
\end{eqnarray}
almost surely.
\end{lemma}

Therefore, applying Lemma \ref{qualem},
we obtain that
\begin{eqnarray}
\label{sinrde3}
{1\over{NL}}{\bf{h}}_k^{\sf{H}}{\bf{V}}^{[k]}\left({1\over{NL}}{\bf{H}}^{[k]}{\bf{P}}^{[k]}{\bf{H}}^{[k]\sf{H}}\right){\bf{V}}{\bf{h}}_k-{\mu'\over{1+{\lambda}_k^{\circ}\mu}}\stackrel{N\rightarrow\infty}{\longrightarrow}0\nonumber\\
\end{eqnarray}  
almost surely, where $\mu={1\over{NL}}{\rm{Tr}}\left({\bf{\Theta}}_k{\bf{V}}^{[k]}\right)$
and $\mu'={1\over{NL}}{\rm{Tr}}\left({1\over{NL}}{\bf{P}}^{[k]}{\bf{H}}^{[k]\sf{H}}{\bf{V}}^{[k]}{\bf{\Theta}}_k{\bf{V}}^{[k]}{\bf{H}}^{[k]}\right)$.
Furthermore, we have
\begin{eqnarray}
\label{sinrde4}
&&{{\lambda}_k^{\circ}\over{NL}}{\bf{h}}_k^{\sf{H}}{\bf{V}}{\bf{h}}_k\left[{1\over{NL}}{\bf{h}}_k^{\sf{H}}{\bf{V}}^{[k]}\left({1\over{NL}}{\bf{H}}^{[k]}{\bf{P}}^{[k]}{\bf{H}}^{[k]\sf{H}}\right){\bf{V}}{\bf{h}}_k\right]-\nonumber\\
&&{{{\lambda}_k^{\circ}\mu\mu'}\over{(1+{\lambda}_k^{\circ}\mu)^2}}\stackrel{N\rightarrow\infty}{\longrightarrow}
0
\end{eqnarray}
almost surely. Based on the results in (\ref{sinrde2}), (\ref{sinrde3}),
(\ref{sinrde4}), we have
\begin{eqnarray}
{1\over{NL}}{\bf{h}}_k^{\sf{H}}{\bf{V}}\left({1\over{NL}}{\bf{H}}^{[k]}{\bf{P}}^{[k]}{\bf{H}}^{[k]\sf{H}}\right){\bf{V}}{\bf{h}}_k-{{\mu'}\over{(1+\lambda_k^{\circ}\mu)^2}}\stackrel{N\rightarrow\infty}{\longrightarrow}
0\nonumber\\
\end{eqnarray}
almost surely. 

From Lemma {\ref{lemlam}}, we have
$\mu-\psi_k^{\circ}\stackrel{N\rightarrow\infty}{\longrightarrow}
0$
almost surely. 

Applying \cite[Lemma 6]{Debbah_TIT2012}, we have
\begin{eqnarray}
&&{1\over{NL}}{\rm{Tr}}\left({1\over{NL}}{\bf{P}}^{[k]}{\bf{H}}^{[k]{\sf{H}}}{\bf{V}}^{[k]}{\bf{\Theta}}_k{\bf{V}}^{[k]}{\bf{H}}^{[k]}\right)-\nonumber\\
&&{1\over{NL}}{\rm{Tr}}\left({1\over{NL}}{\bf{P}}^{[k]}{\bf{H}}^{[k]{\sf{H}}}{\bf{V}}{\bf{\Theta}}_k{\bf{V}}{\bf{H}}^{[k]}\right)\stackrel{N\rightarrow\infty}{\longrightarrow}
0
\end{eqnarray}
almost surely. We further rewrite 
\begin{eqnarray}
&&{1\over{NL}}{\rm{Tr}}\left({1\over{NL}}{\bf{P}}^{[k]}{\bf{H}}^{[k]{\sf{H}}}{\bf{V}}{\bf{\Theta}}_k{\bf{V}}{\bf{H}}^{[k]}\right)\nonumber\\
&&={1\over{NL}}\sum_{i\ne
k} p_i{{{1\over{NL}}{\bf{h}}_i^{\sf{H}}{\bf{V}}{\bf{\Theta}}_k{\bf{V}}{\bf{h}}_i}\over{{1\over{NL}}{\bf{h}}_i^{\sf{H}}{\bf{V}}^2{\bf{h}}_i}}.
\end{eqnarray} 
Applying \cite[Lemma 1, 4 and 6]{Debbah_TIT2012}, we obtain almost surely
\begin{eqnarray}
{{1\over{NL}}{\bf{h}}_i^{\sf{H}}{\bf{V}}{\bf{\Theta}}_k{\bf{V}}{\bf{h}}_i}-{{{1\over{NL}}{\rm{Tr}}({\bf{\Theta}}_i{\bf{V}}{\bf{\Theta}}_k{\bf{V}})}\over{\left[1+{1\over{NL}}{\lambda}_i^{\circ}{\rm{Tr}}({\bf{\Theta}}_i{\bf{V}})\right]^2}}\rightarrow
0.
\end{eqnarray}
To derive a deterministic equivalent for  ${1\over{NL}}{\rm{tr}}\left({\bf{\Theta}}_i{\bf{V}}{\bf{\Theta}}_k{\bf{V}}\right)$,
we write 
\begin{eqnarray}
{1\over{NL}}{\rm{Tr}}\left({\bf{\Theta}}_i{\bf{V}}{\bf{\Theta}}_k{\bf{V}}\right)=\left.
{1\over{NL}}{{\partial}\over{\partial z}}{\rm{Tr}}\left({\bf{\Theta}}_i\left({\bf{V}}^{-1}-z{\bf{\Theta}}_k\right)^{-1}\right)\right|_{z=0}.\nonumber\\
\end{eqnarray}
Observe now that \cite[Theorem 1]{Debbah_TIT2012}
\begin{eqnarray}
{\rm{Tr}}\left({\bf{\Theta}}_i\left({\bf{V}}^{-1}-z{\bf{\Theta}}_k\right)^{-1}\right)-\psi_{ik}(z)\rightarrow0,
\end{eqnarray}
almost surely, where $\psi_{ik}(z)$ is given by $\psi_{ik}(z)={1\over{NL}}{\rm{Tr}}({\bf{\Theta}}_i{\bf{T}}_k(z))$
and ${\bf{T}}_k(z)$ is computed as
\begin{eqnarray}
{\bf{T}}_k(z)=\left({1\over{NL}}\sum_{j=1}^{K}{{{\lambda}_j^{\circ}{\bf{\Theta}}_j}\over{1+{\lambda}_j^{\circ}\psi_{jk}(z)}}+{\bf{Q}}-z{\bf{\Theta}}_k\right)^{-1}.
\end{eqnarray}
By differentiating along $z$, we have $\psi_{ik}'(z)={1\over{NL}}{\rm{Tr}}({\bf{\Theta}}_i{\bf{T}}_k'(z))$,
where ${\bf{T}}_k'(z)={{\partial{\bf{T}}_k(z)}\over{\partial z}}$ is given
by
\begin{eqnarray} 
{\bf{T}}_k'(z)={\bf{T}}_k(z)\left({1\over{NL}}\sum_{j=1}^{K}{{{\lambda}_j^{\circ2}\psi_{jk}'(z){\bf{\Theta}}_j}\over{\left(1+{\lambda}_k^{\circ}\psi_{jk}(z)\right)^2}}+{\bf{\Theta}}_k\right){\bf{T}}_k(z).\nonumber\\
\end{eqnarray}
Setting $z=0$ yields
\begin{eqnarray}
\label{deq2}
\psi_{ik}'(0)={1\over{NL}}{\rm{Tr}}({\bf{\Theta}}_i{\bf{T}}_k'(0)),
\end{eqnarray}
where 
\begin{eqnarray}
\label{deq1}
\!\!\!\!\!\!\!{\bf{T}}_k'(0)={\bf{T}}_k(0)\left({1\over{NL}}\sum_{j=1}^{K}{{{\lambda}_j^{\circ2}\psi_{jk}'(0){\bf{\Theta}}_k}\over{(1+\gamma_k)^2}}+{\bf{\Theta}}_k\right){\bf{T}}_{k}(0).
\end{eqnarray}
In writing the above result, we have taken into account that ${\bf{A}}={\bf{T}}_j(0)$,
$\psi_{jk}(0)={1\over{NL}}{\rm{Tr}}({\bf{\Theta}}_j{\bf{T}}_k(0))={1\over{NL}}{\rm{Tr}}({\bf{\Theta}}_j{\bf{A}})=\psi_j^{\circ}$
and $\gamma_j={\lambda}_j^{\circ}\psi_j^{\circ}$. Plugging (\ref{deq1}) into (\ref{deq2})
and neglecting the functional dependence from $z=0$, ${\bs{\psi}}_k'=[\psi_{1k}'\dots,
\psi_{Kk}']^T$ is found as the unique solution of 
\begin{eqnarray}
\psi_{ik}'&=&{1\over{NL}}{\rm{Tr}}\left({\bf{\Theta}}_i{\bf{A}}\left({1\over{NL}}\sum_{j=1}^{K}{{{\lambda}_j^{\circ2}\psi_{jk}'{\bf{\Theta}}_j}\over{(1+\gamma_j)^2}}+{\bf{\Theta}}_k\right){\bf{A}}\right)\nonumber\\
&=&{1\over{NL}}{\rm{Tr}}({\bf{\Theta}}_i{\bf{A}}{\bf{\Theta}}_k{\bf{A}})+\nonumber\\
&&{1\over{NL}}\sum_{j=1}^{K}{{{\lambda}_j^{\circ2}\psi_{jk}'}\over{(1+\gamma_j)^2}}{1\over{NL}}{\rm{Tr}}\left({\bf{\Theta}}_i{\bf{A}}{\bf{\Theta}}_j{\bf{A}}\right).
\end{eqnarray}

Observing that 
\begin{eqnarray}
{1\over{NL}}{\rm{Tr}}({\bf{\Theta}}_i{\bf{A}}{\bf{\Theta}}_j{\bf{A}})={1\over{L}}\sum_{l=1}^{L}d_{il}d_{jl}\eta_l^2.
\end{eqnarray}

Let ${\bs{\psi}}_k'=[\psi_{1k}',\dots, \psi_{Kk}']^T\in\mathbb{C}^{K}$ and ${\bf{J}}=[J_{ij}]\in\mathbb{C}^{K\times
K}$ with
\begin{eqnarray}
J_{ij}={{{\lambda}_j^{\circ2}}\over{NL(1+\gamma_j)^2}}\left({1\over{L}}\sum_{l=1}^{L}d_{il}d_{jl}\eta_l^2\right).
\end{eqnarray}
We can rewrite the above system of equations in the compact form as 
\begin{eqnarray}
{\bs{\psi}}_k'={\bf{c}}_k+{\bf{J}}{\bs{\psi}}_k',
\end{eqnarray}
where ${\bf{c}}_k=[c_{ik}]$ with $c_{ik}={1\over{L}}\sum_{l=1}^{L}d_{il}d_{jl}\eta_l^2$.

On the basis of above results, using ${1\over{NL}}{\lambda}_i^{\circ}{\rm{tr}}({\bf{\Theta}}_i{\bf{A}})-\gamma_i\rightarrow
0$, we eventually obtain that
\begin{eqnarray}
\label{reseq1}
{1\over{NL}}{\bf{h}}_i^{\sf{H}}{\bf{V}}{\bf{\Theta}}_k{\bf{V}}{\bf{h}}_i-{{\psi_{ik}'}\over{(1+\gamma_i)^2}}\stackrel{N\rightarrow\infty}{\longrightarrow}
0
\end{eqnarray}
almost surely. Following similar arguments of above yields
\begin{eqnarray}
\label{reseq2}
{1\over{NL}}{\bf{h}}_i^{\sf{H}}{\bf{V}}^2{\bf{h}}_i-{{\psi_i'}\over{(1+\gamma_i)^2}}\stackrel{N\rightarrow\infty}{\longrightarrow}
0
\end{eqnarray}
almost surely with ${\bs{\psi}}'=[\psi_1',\dots, \psi_K']^T=({\bf{I}}_K-{\bf{J}})^{-1}{\bf{c}}$
where ${\bf{c}}\in\mathbb{C}^{K}$ has elements $[{\bf{c}}]_i={1\over{L}}\sum_{l=1}^{L}d_{li}\eta_l^2$.

Putting the results in (\ref{reseq1}) and (\ref{reseq2}) together,
we obtain almost surely
\begin{eqnarray}
{1\over{NL}}\sum_{i\ne
k} p_i{{{1\over{NL}}{\bf{h}}_i^{\sf{H}}{\bf{V}}{\bf{\Theta}}_k{\bf{V}}{\bf{h}}_i}\over{{1\over{NL}}{\bf{h}}_i^{\sf{H}}{\bf{V}}^2{\bf{h}}_i}}-{1\over{NL}}\sum_{i\ne
k}p_i{{\psi_{ik}'}\over{\psi_i'}}\stackrel{N\rightarrow\infty}{\longrightarrow} 0.
\end{eqnarray}
The deterministic equivalent of the numerator of the SINR in (\ref{sinrnew})
is now easily obtained as
\begin{eqnarray}
p_k{{|{1\over{NL}}{\bf{h}}_k^{\sf{H}}{\bf{v}}_k|^2}\over{{1\over{NL}}{\bf{h}}_k^{\sf{H}}{\bf{V}}^2{\bf{h}}_k}}-p_k{\psi_k^{\circ2}\over{\psi_k'}}\stackrel{N\rightarrow\infty}{\longrightarrow}
0,
\end{eqnarray}
since ${1\over{NL}}{\bf{h}}_k^{\sf{H}}{\bf{v}}_k-{\psi_k^\circ\over{1+\gamma_k}}\stackrel{N\rightarrow\infty}{\longrightarrow}
0$ and ${1\over{NL}}{\bf{h}}_k^{\sf{H}}{\bf{V}}^2{\bf{h}}_k-{\psi_k'\over{(1+\gamma_k)^2}}\stackrel{N\rightarrow\infty}{\longrightarrow}
0$ almost surely.

Therefore, the deterministic equivalent SINR is given by
\begin{eqnarray}
\label{desinr}
{\sf{sinr}}_k^{\circ}={\psi_k^{\circ2}\over{\psi_k'}}{{p_k}\over{{I}}_k^{\circ}+\sigma_k^2},
\end{eqnarray} 
where 
\begin{eqnarray}
I_k^{\circ}:={1\over{(1+\gamma_k)^2}}\left({1\over{NL}}\sum_{i=1}^{K}p_i{{\psi_{ik}'}\over{\psi_i'}}\right).
\end{eqnarray}

From (\ref{desinr}), it follows that $p_k^{\circ}$ such that ${\sf{sinr}}_k^{\circ}=\gamma_k$
is obtained as
\begin{eqnarray}
p_k^{\circ}= \gamma_k{\psi_k'\over{\psi_k^{\circ2}}}(I_k^{\circ}+\sigma_k^2),
\end{eqnarray}
which can be further rewritten as
\begin{eqnarray}
{p}_k^{\circ}{\psi_k^{\circ2}\over{\psi_k'}}{1\over{\gamma_k}}={I}_k^{\circ}+\sigma_k^2={1\over{(1+\gamma_k)^2}}\left({1\over{NL}}\sum_{i=1}^{K}p_i^{\circ}{{\psi_{ik}'}\over{\psi_i'}}\right)+\sigma_k^2.
\end{eqnarray}
Therefore, the deterministic equivalent for the powers is given by
\begin{displaymath}
\left[ \begin{array}{ccc}
p_1^{\circ}  \\
\vdots \\
p_K^{\circ} 
\end{array} \right]={\bf{M}}^{\circ-1}\left[ \begin{array}{ccc}
\sigma_1^2  \\
\vdots \\
\sigma_K^2 
\end{array} \right],
\end{displaymath}
where
\begin{displaymath}
[{\bf{M}}^{\circ}]_{ij} = \left\{ \begin{array}{ll}
{{1}\over{\gamma _i}}{\psi_i^{\circ2}\over{\psi_i'}}
&, i= j,\\
-{1\over{(1+\gamma_i^2)LN}} {\psi_{ji}'\over{\psi_j'}}
&, i\ne j.
  \end{array} \right.
\end{displaymath} 
We thus complete the proof for the asymptotic results for optimal powers
in Lemma {\ref{detpower}}.

\section{Proof of Theorem \ref{detspa}}
\label{prodetspa}
We rewrite  $\|\tilde{\bf{v}}_l\|_2^2$ as
\begin{eqnarray}
\|\tilde{\bf{v}}_l\|_2^2&=&\sum_{k=1}^{K}{\bf{v}}_k^{\sf{H}}{\bf{Q}}_{lk}{\bf{v}}_k={1\over{NL}}\sum_{k=1}^{K}p_k{{\tilde{\bf{v}}_k^{\sf{H}}{\bf{Q}}_{lk}\tilde{\bf{v}}_k}\over{\|\tilde{\bf{v}}_k\|_2^2}}\nonumber\\
&=&{1\over{NL}}\sum_{k=1}^{K}p_k{{{\bf{h}}_k^{\sf{H}}{\bf{V}}{\bf{Q}}_{lk}{\bf{V}}{\bf{h}}_k}\over{{\bf{h}}_k^{\sf{H}}{\bf{V}}^2{\bf{h}}_k}},
\end{eqnarray}
where ${\bf{Q}}_{lk}\in\mathbb{C}^{NL\times NL}$ is a block diagonal matrix
with the identify matrix ${\bf{I}}_N$ as the $l$-th main diagonal block square
matrix and zeros elsewhere. 

Based on the similar arguments in Appendix {\ref{prodetpower}}, we have the deterministic equivalents 
\begin{eqnarray}
{\bf{h}}_k^{\sf{H}}{\bf{V}}{\bf{Q}}_{lk}{\bf{V}}{\bf{h}}_k-{\psi_{kl}\over{(1+\gamma_k)^2}}\stackrel{N\rightarrow\infty}{\longrightarrow} 0,
\end{eqnarray}
where 
\begin{eqnarray}
\psi_{kl}&=&{1\over{NL}}{\rm{Tr}}({\bf{\Theta}}_k{\bf{A}}{\bf{Q}}_{kl}{\bf{A}})\nonumber\\
&=&{1\over{NL}}\sum_{j=1}^{K}{{{\lambda}_j^{\circ2}\psi_{jl}}\over{(1+\gamma_j)^2}}{1\over{NL}}{\rm{Tr}}\left({\bf{\Theta}}_i{\bf{A}}{\bf{\Theta}}_j{\bf{A}}\right).
\end{eqnarray}
Observing that
\begin{eqnarray}
{1\over{NL}}{\rm{Tr}}({\bf{\Theta}}_k{\bf{A}}{\bf{Q}}_{kl}{\bf{A}})={1\over{L}}d_{kl}\eta_l^2,
\end{eqnarray}
we obtain the final result. We thus complete the proof for the asymptotic results for $\|\tilde{\bf{v}}_l^{[n]}\|_2^2$'s 
in Theorem {\ref{detspa}}.

\bibliographystyle{ieeetr}
\bibliography{/Users/yuanming/OneDrive/Paper/topics/Reference}

\begin{thebibliography}{10}

\bibitem{Yuanming_ISIT16SGSBF}
Y.~Shi, J.~Zhang, and K.~B. Letaief, ``Statistical group sparse beamforming for
  green cloud-ran via large system analysis,'' in {\em Proc. IEEE Int. Symp.
  Inform. Theory (ISIT)}, pp.~870--874, Jul. 2016.

\bibitem{Bhushan_2014networkdensification}
N.~Bhushan, J.~Li, D.~Malladi, R.~Gilmore, D.~Brenner, A.~Damnjanovic,
  R.~Sukhavasi, C.~Patel, and S.~Geirhofer, ``Network densification: the
  dominant theme for wireless evolution into 5{G}.,'' {\em IEEE Commun. Mag.},
  vol.~52, pp.~82--89, Feb. 2014.

\bibitem{Yuanming_WCMLargeCVX}
Y.~Shi, J.~Zhang, K.~Letaief, B.~Bai, and W.~Chen, ``Large-scale convex
  optimization for ultra-dense {C}loud-{RAN},'' {\em IEEE Wireless Commun.
  Mag.}, vol.~22, pp.~84--91, Jun. 2015.

\bibitem{Andrews2015we}
J.~G. Andrews, X.~Zhang, G.~D. Durgin, and A.~K. Gupta, ``Are we approaching
  the fundamental limits of wireless network densification?,'' {\em IEEE
  Commun. Mag.}, vol.~54, pp.~184--190, Oct. 2016.

\bibitem{Jeff_JSAC5G}
J.~Andrews, S.~Buzzi, W.~Choi, S.~Hanly, A.~Lozano, A.~Soong, and J.~Zhang,
  ``What will 5{G} be?,'' {\em IEEE J. Sel. Areas Commun.}, vol.~32,
  pp.~1065--1082, Jun. 2014.

\bibitem{Peng_HCRAN}
M.~Peng, Y.~Li, J.~Jiang, J.~Li, and C.~Wang, ``Heterogeneous cloud radio
  access networks: a new perspective for enhancing spectral and energy
  efficiencies,'' {\em IEEE Wireless Commun. Mag.}, vol.~21, pp.~126--135, Dec.
  2014.

\bibitem{Yuanming_TWC2014}
Y.~Shi, J.~Zhang, and K.~B. Letaief, ``Group sparse beamforming for green
  {C}loud-{RAN},'' {\em IEEE Trans. Wireless Commun.}, vol.~13, pp.~2809--2823,
  May 2014.

\bibitem{Wubben_SPM2014Cloud}
D.~Wubben, P.~Rost, J.~Bartelt, M.~Lalam, V.~Savin, M.~Gorgoglione, A.~Dekorsy,
  and G.~Fettweis, ``Benefits and impact of cloud computing on {5G} signal
  processing: Flexible centralization through {C}loud-{RAN},'' {\em IEEE Signal
  Process. Mag.}, vol.~31, pp.~35--44, Nov. 2014.

\bibitem{Gesbert_JSAC10}
D.~Gesbert, S.~Hanly, H.~Huang, S.~Shamai~Shitz, O.~Simeone, and W.~Yu,
  ``Multi-cell {MIMO} cooperative networks: A new look at interference,'' {\em
  IEEE J. Sel. Areas Commun.}, vol.~28, pp.~1380--1408, Sep. 2010.

\bibitem{Peng_2015fronthaul}
M.~Peng, C.~Wang, V.~Lau, and H.~V. Poor, ``Fronthaul-constrained cloud radio
  access networks: Insights and challenges,'' {\em IEEE Wireless Commun.},
  vol.~22, pp.~152--160, Apr. 2015.

\bibitem{Shamai_SPM2014Fronthaul}
S.~Park, O.~Simeone, O.~Sahin, and S.~Shamai~Shitz, ``Fronthaul compression for
  cloud radio access networks: Signal processing advances inspired by network
  information theory,'' {\em IEEE Signal Process. Mag.}, vol.~31, pp.~69--79,
  Nov. 2014.

\bibitem{leyffer_2012mixed}
S.~Leyffer, {\em Mixed integer nonlinear programming}, vol.~154.
\newblock Springer, 2012.

\bibitem{Boyd_2008enhancing}
E.~J. Candes, M.~B. Wakin, and S.~P. Boyd, ``Enhancing sparsity by reweighted
  $\ell_1$ minimization,'' {\em J. Fourier Anal. Appl.}, vol.~14, pp.~877--905,
  Dec. 2008.

\bibitem{Daubechies_2010iteratively}
I.~Daubechies, R.~DeVore, M.~Fornasier, and C.~S. G{\"u}nt{\"u}rk,
  ``Iteratively reweighted least squares minimization for sparse recovery,''
  {\em Communications on Pure and Applied Mathematics}, vol.~63, no.~1,
  pp.~1--38, 2010.

\bibitem{Yuanming_JSAC2015}
Y.~Shi, J.~Cheng, J.~Zhang, B.~Bai, W.~Chen, and K.~B. Letaief, ``Smoothed
  ${L}_p$-minimization for green {C}loud-{RAN} with user admission control,''
  {\em IEEE J. Sel. Areas Commun.}, vol.~34, pp.~1022--1036, Apr. 2016.

\bibitem{Yuanming_LargeSOCP2014}
Y.~Shi, J.~Zhang, B.~O'Donoghue, and K.~Letaief, ``Large-scale convex
  optimization for dense wireless cooperative networks,'' {\em IEEE Trans.
  Signal Process.}, vol.~63, pp.~4729--4743, Sept. 2015.

\bibitem{boyd2011distributed}
S.~Boyd, N.~Parikh, E.~Chu, B.~Peleato, and J.~Eckstein, ``Distributed
  optimization and statistical learning via the alternating direction method of
  multipliers,'' {\em Found. Trends in Mach. Learn.}, vol.~3, pp.~1--122, Jul.
  2011.

\bibitem{Boyd_arXiv2013}
B.~O'Donoghue, E.~Chu, N.~Parikh, and S.~Boyd, ``Conic optimization via
  operator splitting and homogeneous self-dual embedding,'' {\em J. Optim.
  Theory Appl.}, pp.~1--27, Feb. 2016.

\bibitem{hunter2004tutorial}
D.~R. Hunter and K.~Lange, ``A tutorial on {MM} algorithms,'' {\em Amer.
  Statistician}, vol.~58, no.~1, pp.~30--37, 2004.

\bibitem{Yu_2007transmitter}
W.~Yu and T.~Lan, ``Transmitter optimization for the multi-antenna downlink
  with per-antenna power constraints,'' {\em IEEE Trans. Signal Process.},
  vol.~55, no.~6, pp.~2646--2660, 2007.

\bibitem{Wei_IA2014Sparse}
B.~Dai and W.~Yu, ``Sparse beamforming and user-centric clustering for downlink
  cloud radio access network,'' {\em IEEE Access}, vol.~2, pp.~1326--1339, Nov.
  2014.

\bibitem{Verdu_RMT04}
A.~M. Tulino and S.~Verdu, ``Random matrix theory and wireless
  communications,'' {\em Found. Trends in Commun. and Inf. Theory}, vol.~1,
  no.~1, pp.~1--182, 2004.

\bibitem{Couillet_2011random}
R.~Couillet and M.~Debbah, {\em Random matrix methods for wireless
  communications}.
\newblock Cambridge University Press, 2011.

\bibitem{Debbah_TIT2012}
S.~Wagner, R.~Couillet, M.~Debbah, and D.~T.~M. Slock, ``Large system analysis
  of linear precoding in correlated {MISO} broadcast channels under limited
  feedback,'' {\em IEEE Trans. Inf. Theory}, vol.~58, pp.~4509--4537, Jul.
  2012.

\bibitem{Hanly_2012TIT}
R.~Zakhour and S.~V. Hanly, ``Base station cooperation on the downlink: Large
  system analysis,'' {\em IEEE Trans. Inf. Theory}, vol.~58, pp.~2079--2106,
  Apr. 2012.

\bibitem{Couillet_TWC16}
L.~Sanguinetti, R.~Couillet, and M.~Debbah, ``Large system analysis of base
  station cooperation for power minimization,'' {\em IEEE Trans. Wireless
  Commun.}, vol.~15, pp.~5480--5496, Aug. 2016.

\bibitem{Couillet_2014large}
R.~Couillet and M.~McKay, ``Large dimensional analysis and optimization of
  robust shrinkage covariance matrix estimators,'' {\em Journal of Multivariate
  Analysis}, vol.~131, pp.~99--120, 2014.

\bibitem{TonyQ.S._WC2013}
J.~Zhao, T.~Q. Quek, and Z.~Lei, ``Coordinated multipoint transmission with
  limited backhaul data transfer,'' {\em IEEE Trans. Wireless Commun.},
  vol.~12, pp.~2762--2775, Jun. 2013.

\bibitem{Z.Q.Luo_JSAC2013}
M.~Hong, R.~Sun, H.~Baligh, and Z.-Q. Luo, ``Joint base station clustering and
  beamformer design for partial coordinated transmission in heterogeneous
  networks,'' {\em IEEE J. Sel. Areas Commun.}, vol.~31, pp.~226--240, Feb.
  2013.

\bibitem{Rui_TWC2015GSBF}
S.~Luo, R.~Zhang, and T.~J. Lim, ``Downlink and uplink energy minimization
  through user association and beamforming in {C}-{RAN},'' {\em IEEE Trans.
  Wireless Commun.}, vol.~14, pp.~494--508, Jan. 2015.

\bibitem{Wei_JSAC16}
B.~Dai and W.~Yu, ``Energy efficiency of downlink transmission strategies for
  cloud radio access networks,'' {\em IEEE J. Sel. Areas Commun.}, vol.~34,
  pp.~1037--1050, Apr. 2016.

\bibitem{Yuanming_SDP2014}
Y.~Shi, J.~Zhang, and K.~Letaief, ``Robust group sparse beamforming for
  multicast green {C}loud-{RAN} with imperfect {CSI},'' {\em IEEE Trans. Signal
  Process.}, vol.~63, pp.~4647--4659, Sept. 2015.

\bibitem{Jun_caching2014}
X.~Peng, J.-C. Shen, J.~Zhang, and K.~B. Letaief, ``Joint data assignment and
  beamforming for backhaul limited caching networks,'' in {\em Proc. IEEE Int.
  Symp. on Personal Indoor and Mobile Radio Comm. (PIMRC)}, (Washington, DC),
  Sep. 2014.

\bibitem{Tao_TWC16}
M.~Tao, E.~Chen, H.~Zhou, and W.~Yu, ``Content-centric sparse multicast
  beamforming for cache-enabled {C}loud {RAN},'' {\em IEEE Trans. Wireless
  Commun.}, vol.~15, pp.~6118--6131, Sep. 2016.

\bibitem{Tony_2015heterogeneous}
J.~Zhao, T.~Q. Quek, and Z.~Lei, ``Heterogeneous cellular networks using
  wireless backhaul: Fast admission control and large system analysis,'' {\em
  IEEE J. Sel. Areas Commun.}, pp.~2128--2143, Oct. 2015.

\bibitem{Yuanming_Compoffloading}
J.~Cheng, Y.~Shi, B.~Bai, and W.~Chen, ``Computation offloading in cloud-ran
  based mobile cloud computing system,'' in {\em IEEE Int. Conf. on Commun.
  (ICC), Kuala Lumpur, Malaysia}, pp.~1--6, May 2016.

\bibitem{Luo_2013base}
W.-C. Liao, M.~Hong, Y.-F. Liu, and Z.-Q. Luo, ``Base station activation and
  linear transceiver design for optimal resource management in heterogeneous
  networks,'' {\em IEEE Trans. Signal Process.}, vol.~62, pp.~3939--3952, Aug.
  2014.

\bibitem{Rusek_SPM2013}
F.~Rusek, D.~Persson, B.~K. Lau, E.~Larsson, T.~Marzetta, O.~Edfors, and
  F.~Tufvesson, ``Scaling up {MIMO}: Opportunities and challenges with very
  large arrays,'' {\em IEEE Signal Process. Mag.}, vol.~30, pp.~40--60, Jan.
  2013.

\bibitem{Couillet_SPM13LS}
R.~Couillet and M.~Debbah, ``Signal processing in large systems: A new
  paradigm,'' {\em IEEE Signal Process. Mag.}, vol.~30, pp.~24--39, Jan. 2013.

\bibitem{Tropp_15RMT}
J.~A. Tropp, ``An introduction to matrix concentration inequalities,'' {\em
  Found. Trends in Mach. Learn.}, vol.~8, pp.~1--230, May 2015.

\bibitem{Couillet_2016randomICML}
R.~Couillet, G.~Wainrib, H.~T. Ali, and H.~Sevi, ``A random matrix approach to
  echo-state neural networks,'' in {\em Proc. Int. Conf. Mach. Learn. (ICML),
  33th}, pp.~517--525, 2016.

\bibitem{Ding_CMagBigData}
S.~Bi, R.~Zhang, Z.~Ding, and S.~Cui, ``Wireless communications in the era of
  big data,'' {\em IEEE Commun. Mag.}, vol.~53, pp.~190--199, Oct. 2015.

\bibitem{Debbah_MMIMOTIT2015}
S.~Lakshminarayana, M.~Assaad, and M.~Debbah, ``Coordinated multicell
  beamforming for massive {MIMO}: A random matrix approach,'' {\em IEEE Trans.
  Inf. Theory}, vol.~61, pp.~3387--3412, Jun. 2015.

\bibitem{Lange_1993normal}
K.~Lange and J.~S. Sinsheimer, ``Normal/independent distributions and their
  applications in robust regression,'' {\em J. Comput. Graph. Stat.}, vol.~2,
  no.~2, pp.~175--198, 1993.

\bibitem{WeiYu_WC10}
H.~Dahrouj and W.~Yu, ``Coordinated beamforming for the multicell multi-antenna
  wireless system,'' {\em IEEE Trans. Wireless Commun.}, vol.~9,
  pp.~1748--1759, Sep. 2010.

\bibitem{boyd2004convex}
S.~P. Boyd and L.~Vandenberghe, {\em Convex optimization}.
\newblock Cambridge University Press, 2004.

\bibitem{Rockafellar1997convex}
R.~T. Rockafellar, {\em Convex analysis}, vol.~28.
\newblock Princeton university press, 1997.

\bibitem{horn2012matrix}
R.~A. Horn and C.~R. Johnson, {\em Matrix analysis}.
\newblock Cambridge university press, 2012.

\end{thebibliography}

\end{document}